\documentclass[aps,preprint,nofootinbib,preprintnumbers,eqsecnum]{revtex4-1}

\usepackage[usenames, dvipsnames]{color}

\newcommand{\nb}[1]{\color{blue}}

\usepackage[
	  pagebackref=false,
	  colorlinks=true,
      linkcolor=blue,
      urlcolor=blue,
      filecolor=black,
      citecolor=red,
      pdfstartview=FitV,
      pdftitle={},
        pdfauthor={},
        pdfsubject={},
        pdfkeywords={},
        pdfpagemode=None,
        bookmarksopen=true
      ]{hyperref}

\usepackage[normalem]{ulem}
\usepackage{amsmath}
\usepackage{enumerate}
\usepackage{amsfonts}
\usepackage{amssymb}
\usepackage{epsfig}
\usepackage{mathbbol}

\def\Tr{\mathop{\rm Tr}}
\def\tr{\mathop{\rm tr}}

\newcommand\half{{\ensuremath{\frac{1}{2}}}}
\newcommand\p{\ensuremath{\partial}}

\newcommand\field[1]{{\ensuremath{\mathbb{{#1}}}}}

\newcommand\vev[1]{{\ensuremath{\left\langle{#1}\right\rangle}}}

\newcommand\ket[1]{\ensuremath{\lvert{#1}\rangle}}
\newcommand\bra[1]{\ensuremath{\langle{#1}\rvert}}

\newcommand{\RR}{\field{R}}

\newcommand{\be}{\begin{equation}}
\newcommand{\ee}{\end{equation}}
\newcommand{\bea}{\begin{eqnarray}}
\newcommand{\eea}{\end{eqnarray}}
\newcommand{\bega}{\begin{gather}}
\newcommand{\eega}{\end{gather}}

\newcommand{\bi}{\begin{itemize}}
\newcommand{\ei}{\end{itemize}}
\newcommand{\ben}{\begin{enumerate}}
\newcommand{\een}{\end{enumerate}}
\newcommand{\bca}{\begin{cases}}
\newcommand{\eca}{\end{cases}}
\newcommand{\bln}{\begin{align}}
\newcommand{\eln}{\end{align}}
\newcommand{\bst}{\begin{split}}
\newcommand{\est}{\end{split}}
\def\ie{\begin{equation}\begin{aligned}}
\def\fe{\end{aligned}\end{equation}}
\newcommand{\bma}{\le(\begin{matrix}}
\newcommand{\ema}{\end{matrix}\ri)}

\def\b{{\beta}}
\newcommand\ep{\epsilon}

\newcommand\Om{\Omega}

\newcommand\De{{\ensuremath{{\Delta}}}}

\newcommand\da{{\dagger}}

\newcommand\ov{\over}
\newcommand\ha{{\half}}

\def\le{\left}
\def\ri{\right}

\newcommand\sA{{\ensuremath{{\mathcal A}}}}
\newcommand\sB{{\ensuremath{{\mathcal B}}}}

\newcommand\sH{{\ensuremath{{\mathcal H}}}}

\newcommand\sL{{\ensuremath{{\mathcal L}}}}
\newcommand\sM{{\ensuremath{{\mathcal M}}}}
\newcommand\sN{{\ensuremath{{\mathcal N}}}}
\newcommand\sO{{\ensuremath{{\mathcal O}}}}

\newcommand\sS{{\mathcal S}}

\newcommand\sX{{\mathcal X}}
\newcommand\sY{{\mathcal Y}}

\newcommand\bA{{\bf A}}
\newcommand\bB{{\bf B}}

\newcommand\bD{{\bf D}}
\newcommand\bL{{\bf L}}
\newcommand\bR{{\bf R}}

\newcommand\vk{{\vec k}}

\newcommand{\fa}{{\mathfrak a}}
\newcommand{\fb}{{\mathfrak{b}}}
\newcommand{\fc}{{\mathfrak{c}}}

\newcommand{\ft}{{\mathfrak t}}

\newcommand{\fw}{{\mathfrak w}}

\newcommand{\vol}{{\rm Vol}}

\newcommand{\pt}{\partial}

\newcommand{\wt}{\widetilde}

\DeclareMathOperator\arctanh{arctanh}

\usepackage{graphicx}
\usepackage{caption}
\usepackage{subcaption}
\usepackage{mwe}

\begin{document}

\title{Volume as an index of a subalgebra}

\preprint{MIT-CTP/5897}

\author{Sam Leutheusser}
\affiliation{ School of Natural Sciences, Institute for Advanced Study,
Princeton, NJ 08540, USA }

\author{Hong Liu}
\affiliation{Center for Theoretical Physics---a Leinweber Institute, 
Massachusetts
Institute of Technology, \\
77 Massachusetts Ave.,  Cambridge, MA 02139 }

\begin{abstract}
\noindent

We propose a new way to understand the volume of certain {sub}regions in the bulk of AdS spacetime by relating it to an algebraic quantity known as the index of inclusion. This index heuristically measures the relative size of a subalgebra $\sN$ embedded within a larger algebra $\sM$. According to subregion-subalgebra duality, bulk {sub}regions are described by von Neumann algebras on the boundary. When a causally complete bulk {sub}region corresponds to the relative commutant $\sN' \cap \sM$---the set of operators in $\sM$ that commute with $\sN$---{of boundary subalgebras, }we propose that the exponential of the volume of the maximal volume slice of the {sub}region equals the index of inclusion.
This ``volume-index'' relation provides a new boundary explanation for the growth of interior volume in black holes, reframing it as a change in the relative size of operator algebras. It offers a complementary perspective on complexity growth from the Heisenberg picture, and has a variety of other applications, including quantifying the relative size of algebras dual to the entanglement wedge and the causal wedge of a boundary region, as well as quantifying the violation of additivity of operator  algebras in the large $N$ limit. {Finally, it may offer insights into the volume growth of de Sitter space through the changes in North and South pole observer algebras in time.}

\end{abstract}

\today

\maketitle

\tableofcontents

\section{Introduction}

Since the discovery of Bekenstein-Hawking entropy~\cite{Bekenstein:1972tm, Hawking:1975vcx}, the area of spacetime regions in gravitational systems has been a subject of long-standing fascination. This relationship between area and entropy has played a fundamental role in formulating the holographic principle~\cite{tHooft:1993dmi, Susskind:1994vu} and uncovering deep connections between spacetime geometry and quantum entanglement~\cite{Ryu:2006bv, Hubeny:2007xt, Engelhardt:2014gca}. 

In contrast, volume has played a far less starring role---until recently. In the ``complexity = volume'' proposal~\cite{Susskind:2014rva, Stanford:2014jda} it has been postulated that  the complexity of a quantum state in the boundary theory is equal 
to the volume of a maximal spatial slice in the AdS space that extends from the boundary at a given time.\footnote{
The maximal volume slice is a spacelike surface in the AdS space that has the largest possible volume for a given boundary time.}  More explicitly,\footnote{In~\cite{Stanford:2014jda}, equation~\eqref{volcom} was written as an equality, with the understanding that the equation is true up to an ambiguous order one constant. 
}
\be\label{volcom}
C_V \sim {V \ov G_N l_{\rm AdS}}, 
\ee
where $C_V$  is the complexity of the boundary state,
$V$ is the maximal spatial volume in the bulk, $G_N$ is Newton's constant, and 
$l_{\rm AdS}$ is some characteristic AdS curvature radius. {An alternative boundary interpretation of the volume of an entire bulk Cauchy slice has been given in terms of the fidelity susceptibility of the CFT under a marginal perturbation~\cite{Miyaji:2015woj}.}

As the quantum state on the boundary evolves, the complexity of the state increases, and this increase is captured  by the growth of the volume of the maximal slice in the bulk. 
The most compelling support for the conjecture is the linear growth of black hole interior volume and the growth of complexity in chaotic systems.

There have also been proposals to generalize the conjecture to subregions. For example, based on divergence structures, the proposal of~\cite{Alishahiha:2015rta} suggests that the volume of the maximal slice of an entanglement wedge is dual to the reduced fidelity susceptibility for the state on the dual boundary subregion.

Nevertheless, there are significant challenges for the proposal~\eqref{volcom}, including: (i) lack of a precise and unambiguous definition of the complexity in quantum field theory~\cite{Jefferson:2017sdb, Chapman:2017rqy}; (ii) ambiguities in the choice of bulk dual quantity~\cite{Belin:2021bga, Belin:2022xmt, Jorstad:2023kmq, Myers:2024vve}; (iii) lack of a rigorous argument for the conjecture and concrete computations of complexity in the boundary field theory; (iv) lack of a boundary interpretation of divergences from the asymptotic region.\footnote{This is related to the lack of good definition of complexity in QFT. Heuristically, complexity is not a UV sensitive quantity since all finite energy states approach the vacuum in the UV; however, the proposed bulk duals are IR divergent, suggesting, according the the complexity=volume conjecture, that complexity is UV divergent. The vacuum of a QFT will have UV divergent complexity with respect to a local product state; however, there is currently no reason to believe that a local product state is the `correct' choice of reference state for the  complexity of a QFT.}

In this paper, we propose an alternative boundary interpretation of bulk {subregion} volume, as a characterization of the relative size of certain subalgebras. As we will see below, this new interpretation provides a complementary {Heisenberg picture perspective to}
complexity growth.

Our starting point is the subregion-subalgebra duality proposed in~\cite{longPaper, subalgSubreg}, where 
a causally complete bulk subregion $\bB$ is described by an emergent type III$_1$ von Neumann algebra $\sM_\bB$ on the boundary, in the $G_N \to 0$ limit. Heuristically, we would like to argue that the volume of $\fb$, the Cauchy slice of $\bB$ of maximal volume, provides a measure of the ``size'' of $\sM_\bB$. While the size of an infinite dimensional von Neumann algebra is not well-defined, there is a  way to quantify relative size of two algebras if one is a subalgebra of the other. Consider a von Neumann algebra $\sM$ and a subalgebra $\sN \subseteq \sM$. By virtue of the inclusion, $\sM$ is ``larger''  than $\sN$, and this difference can be quantified using an index for the inclusion $\sN \subseteq \sM,$ which we will denote by $[\sM : \sN] \in \mathbb{R}_+ \cup \{\infty\}$.

We propose
{that} the exponential of the bulk volume
{of} certain regions {defines an index for certain boundary algebra inclusions}. Before stating this proposal explicitly, we begin by reviewing the definition of an index of inclusion in a simple setting to build some intuition.

The index $[\sM : \sN]$ is particularly simple to describe for finite type I inclusions. Consider a type I$_n$ von Neumann factor\footnote{A type $I_n$ von Neumann factor is simply the matrix algebra of $n \times n$ matrices.} $\sN$ contained in a type I$_m$ factor $\sM$, where $m > n$. For this to happen, their ratio $k \equiv m/n$ must be a natural number. We can define the index of inclusion of $\sN \subset \sM$ by the number of times $\sN$ can be ``fit into'' $\sM$. 
More explicitly, we can represent $\sM$ and $\sN$ on a $m$-dimensional Hilbert space $\sH_{m} = \sH_{k} \otimes \sH_{n}$ 
as $\sM = \sB(\sH_{m})$ and $\sN = \mathbf{1}_k \otimes \sB(\sH_n)$. 
Collections of operators of the form $\ket{i}\bra{j} \otimes a$---with $a \in \sN$ and $\{\ket{i} \, | \, i=1,2, \cdots, k\}$  being an 
orthonormal basis  for $\sH_k$---give independent copies of $\sN$ for each choice of $(i,j)$.  
There are thus $k^2$ independent copies of $\sN$ that can be fit into one copy of $\sM$, and the index of inclusion for 
$\sN \subset \sM$ is given by $[\sM : \sN] = k^2$.

Another useful (and equivalent) characterization of the index is through the relative commutant. The relative commutant of $\sN$ in $\sM$ is $\sN' \cap \sM \simeq \sB(\sH_k) \otimes \mathbf{1}_n$, which has vector space dimension $k^2$. 
Thus, in this case, the index of inclusion can also be thought of as the ``size'' of the relative commutant.

As a simple example in a familiar physical setting, consider a system of qudits of local Hilbert space dimension $q$ arranged on a cubic (spatial) lattice $L$ in $d$-dimensions. Suppose each side of the lattice has $m$ sites and we take $\sM$ to be the full operator algebra of the system.  
Now consider the subalgebra $\sN$ corresponding to the sublattice $N \subset L$ obtained by keeping the first $n~(< m)$ sites in each direction. Clearly we have $\sN \subset \sM$ and the relative commutant  is generated by qudits lying in the complement $K = L\backslash N$, i.e., $\sN' \cap \sM = \sB (\sH_K)$, which gives 
\be 
	[\sM : \sN] = q^{2(M^d - N^d)} = \exp \le( 2 \log q {{\rm Vol} (K) \ov a^d} \ri)  ,
\ee
where ${\rm Vol (K)}$ denotes the volume of $K$ and $a$ is the lattice spacing. 
In this case, the expression can also be written as
\be 
	[\sM : \sN] = \exp \le( 2 \log q {{\rm Vol} (L) - {\rm Vol} (N) \ov a^d} \ri)  .
\ee

An index $[\sM : \sN]$ can also be defined for an inclusion, $\sN \subseteq \sM,$ of Type I$_\infty$, Type II, and Type III factors. {The most general notion of index was defined by Kosaki~\cite{kosakiIndex} with respect to a conditional expectation $E: \sM \to \sN$. The Kosaki index includes the well-known Jones index~\cite{jonesIndex} for Type II$_1$ inclusions and the Longo index~\cite{longoIndex} for type III inclusions as subcases}. The definitions of these indices require more elaborate mathematical machinery, which is reviewed in Appendix~\ref{app:index}. Nevertheless, the intuition that the index can be thought of as 
a measure of the relative commutant $\sN' \cap \sM$ still holds. Since the Kosaki index can be used to define an index for all types of algebras, below we will collectively refer to these known indices as the Kosaki index.

Now consider boundary algebras $\sN \subseteq \sM$ with the relative commutant $\sN' \cap \sM$ dual to some bulk {domain of dependence} $\bB$. Our proposal is that the maximal volume slice, $\fb$, of $\bB$, defines an index for the embedding of $\sN$ into $\sM$ through
\be \label{v1}
(\sM:\sN) = \exp \le(C ~{\rm Vol} (\fb) \ri) \ .
\ee
Here $C \propto {1 \ov G_N}$ is a dimensional constant.
Equation~\eqref{v1} should be understood as giving the leading order contribution to $\log (\sM:\sN)$ at the leading order in the $1/N$ expansion, which we will elaborate more explicitly in Sec.~\ref{sec:finiteN}. {See~\cite{vanderHeijden:2024tdk} for a distinct recent discussion of the Jones index in a gravitational context.}

\begin{figure}[h!]
        \centering
		\includegraphics[width=0.5\textwidth]{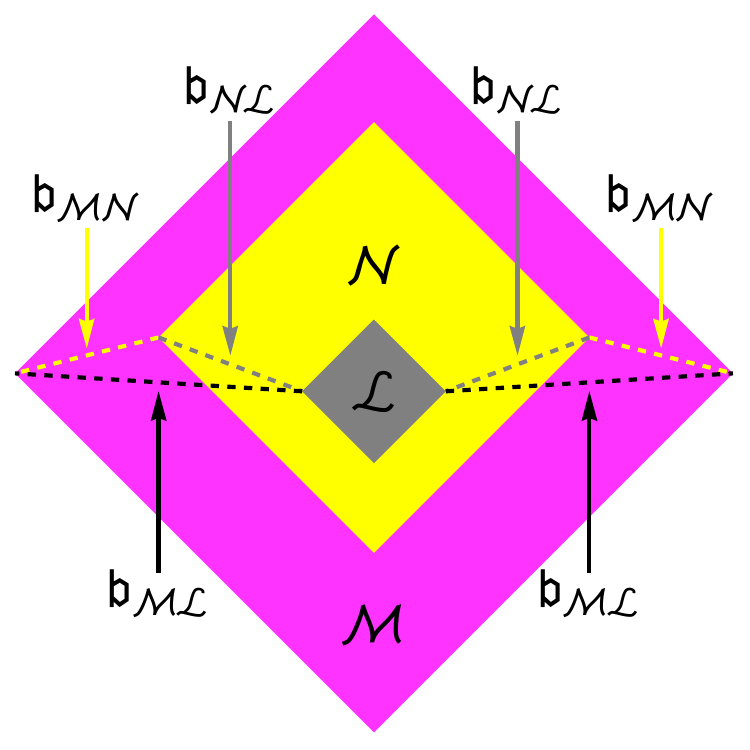}
        \caption[  ]
        {\small The bulk duals of $\sL \subset \sN \subset \sM$ are depicted in gray, yellow, and magenta, respectively. The dashed black line is the maximal volume slice, $\fb_{\sM\sL}$, of bulk dual of $\sL' \cap \sM$. The dashed yellow line is the maximal volume slice, $\fb_{\sM\sN}$, of bulk dual of $\sN' \cap \sM$. The dashed gray line is the maximal volume slice, $\fb_{\sN\sL}$, of bulk dual of $\sL' \cap \sN$. Note that $\fb_{\sM\sL}$ is not the disjoint union of $\fb_{\sM\sN}$ and $\fb_{\sN\sL}$.
        In fact, we have ${\rm Vol}(\fb_{\sM\sL}) \neq {\rm Vol}(\fb_{\sM\sN}) + {\rm Vol}(\fb_{\sN\sL})$ and thus, by~\eqref{v1}, we have $(\sM : \sL) \neq (\sM : \sN) (\sN : \sL)$. However, $\fb_{\sM\sN} \cup \fb_{\sN\sL}$ is a Cauchy slice of the dual of $\sL' \cap \sM$ and thus, since $\fb_{\sM\sL}$ has maximal volume over all such slices we have the inequality ${\rm Vol}(\fb_{\sM\sL}) \geq {\rm Vol}(\fb_{\sM\sN}) + {\rm Vol}(\fb_{\sN\sL})$ leading to~\eqref{inmul} via~\eqref{v1}.  
      } 
\label{fig:notKosaki}
\end{figure}

Currently, it is not clear if~\eqref{v1} can be related to the Kosaki index. Instead, we may view~\eqref{v1} as a definition 
of a new index  using holography. An indication that in general it does not agree with the Kosaki index is that {it does not obey all the properties obeyed by Kosaki's index. For example,}
Kosaki's index is multiplicative. Suppose we have a chain of subfactors inclusions $\sL \subset \sN \subset \sM$, then 
\be \label{multica} 
[\sM: \sL] = [\sM: \sN] [\sN: \sL] \ .
\ee
In contrast, from Fig.~\ref{fig:notKosaki}, we see that the index~\eqref{v1} is in general not multiplicative, but instead satisfies
\be \label{inmul}
(\sM: \sL) \geq (\sM: \sN ) (\sN: \sL) \ .
\ee
The inequality~\eqref{inmul} is saturated, i.e., satisfies~\eqref{multica}, when the maximal volume slices for the regions dual to $\sL'\cap \sM$, $\sN' \cap \sM$ and $\sL' \cap \sN$ all lie on the same bulk Cauchy slice.

\begin{figure}[h!]
        \centering
        \begin{subfigure}[b]{0.55\textwidth}
            \centering
		\includegraphics[width=\textwidth]{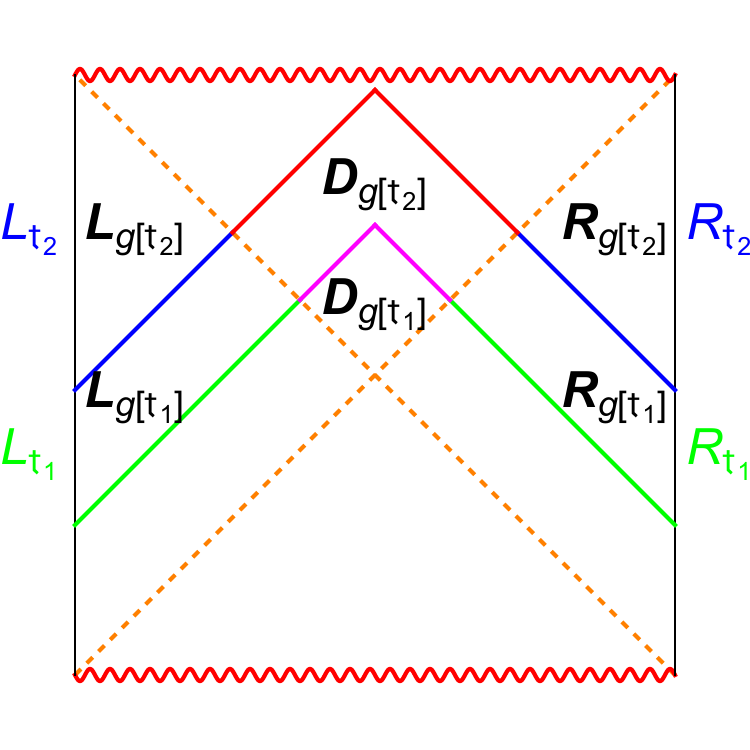} 
            \caption[]%
            {{\small}  Duals of semi-infinite time bands $t \in (\ft_1, \infty)$ (green) and $t \in (\ft_2, \infty)$ (blue). The duals of the commutants of their joint algebras are respectively shown in magenta and red.
           }    
        \end{subfigure}
        \hfill
        \begin{subfigure}[b]{0.35\textwidth}   
            \centering 
		\includegraphics[width=\textwidth]{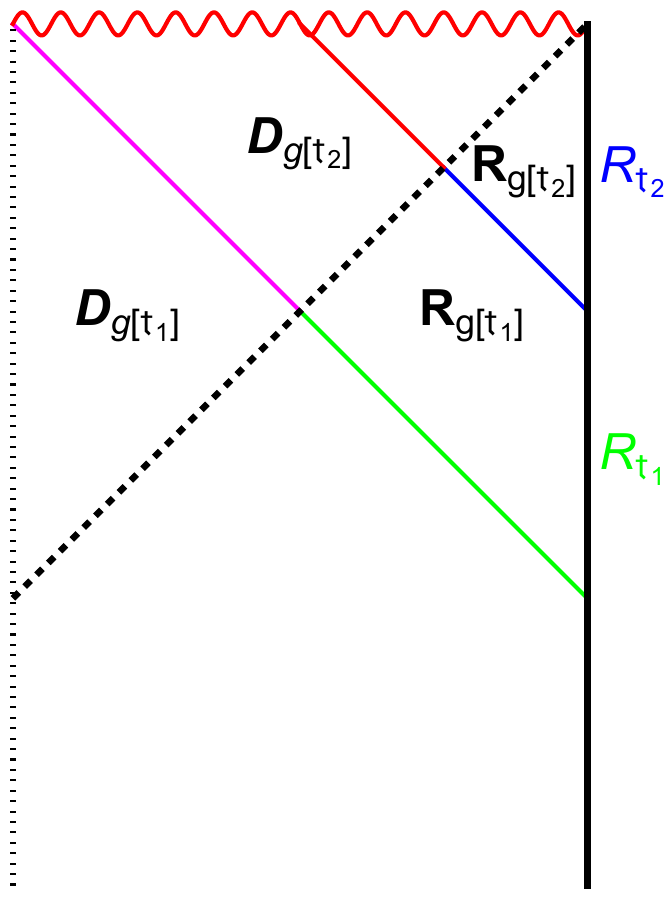}
            \caption[]%
            {{\small   Duals of semi-infinite time bands $t \in (\ft_1, \infty)$ (green) and $t \in (\ft_2, \infty)$ (blue). The duals of the commutants of their joint algebras are respectively shown in magenta and red.
            }}    
        \end{subfigure}
        \caption[  ]
        {\small  The algebra generated by single-trace operators in the semi-infinite time-band $L_{\ft}$ $(R_{\ft})$ is dual to the bulk subregion $\bL_{g(\ft)}$ $(\bR_{g(\ft)})$ that is the exterior of a finite left (right) future horizon cut at affine parameter $g(\ft) > 0.$ In the figures $\ft_2 > \ft_1$ and we see that the dual of the relative commutant algebra has grown, i.e. $\bD_{g(\ft_1)} \subset \bD_{g(\ft_2)} \Leftrightarrow \sN_{\ft_1}' \cap \sM \subset  \sN_{\ft_2}' \cap \sM$.        } 
\label{fig:extDualTimeBands}
\end{figure}

Equation~\eqref{v1} can be used to give a boundary explanation of the interior volume growth of a black hole. 
Consider two copies of a holographic CFT on $\mathbb{R} \times S^{d-1}$ entangled in the TFD state above the Hawking-Page temperature. We define $L_{\ft}~(R_{\ft})$ to be the spacetime subregion of the left (right) boundary with $t > \ft,$ i.e., it is a semi-infinite time-band on the left (right) boundary. We denote the algebras generated by single-trace operators in these subregions by $\sY_{L_{\ft}}$ and $\sY_{R_{\ft}}$, respectively. These algebras are dual to the bulk subregions exterior to finite cuts of the left and right future horizons of the eternal AdS black hole. See Fig.~\ref{fig:extDualTimeBands} (a).

We now take $\sM$ to be the full algebra of the two CFTs and $\sN_\ft = \sY_{L_\ft} \lor \sY_{R_\ft}$. 
The relative commutant $\sN'_\ft \cap \sM$ is then dual to the bulk region $\bD_{g(\ft)}$, 
which is the subregion `in-between' the finite cuts of the left and right future horizons at $g(\ft).$ See Fig.~\ref{fig:extDualTimeBands} (a). Evolving both copies of the CFTs forward in time by some amount $s > 0$ sends the algebras $\sY_{L_{\ft}}/\sY_{R_{\ft}}$ to $\sY_{L_{\ft+s}}/\sY_{R_{\ft+s}}$ which are (proper) subalgebras of the original algebras $\sY_{L_{\ft}}/\sY_{R_{\ft}}$. In equations, we have
\bega 
\sN_{\ft +s} = \sY_{L_{\ft +s}} \vee \sY_{R_{\ft +s}} \subset \sN_\ft = \sY_{L_{\ft}} \vee \sY_{R_{\ft}} \cr
\Rightarrow \quad \sN_\ft' \cap \sM \subset  \sN_{\ft +s}' \cap \sM \ .
\end{gather} 
The growth of the algebra $\sN_{\ft +s}' \cap \sM$ under time evolution then gives a boundary explanation of  
the growth of the region $\bD_{g(\ft +s)}$ as well as the growth of the interior volume of an eternal black hole. 

Taking the converse of this phenomenon, we see that the bulk volume increase of $\bD_{g(\ft+s)},$ dual to the growth of $\sN_{\ft +s}' \cap \sM,$ quantifies how the algebra $\sN_{\ft + s}$ shrinks within $\sM$. 
 
The same argument can be immediately extended  to explain the volume growth of the interior of  
an evaporating black hole, see Fig.~\ref{fig:extDualTimeBands} (b).

\begin{figure}[h]
        \centering
        \includegraphics[width=6cm]{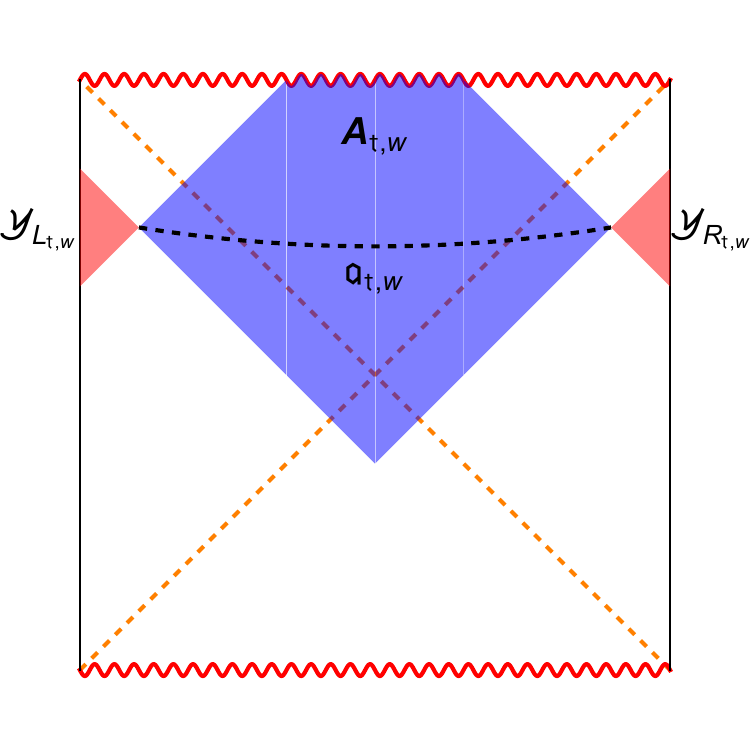}
                \caption[  ]
        {\small The bulk dual of $\sN_{\ft, w}' \cap \sM$ is $\bA_{\ft, w}$ (shown in blue) and the dual of its commutant $\sN_{\ft, w}$ (shown in red). As $\ft$ increases (with $w$ fixed), $\vol \le(\fa_{\ft, w}\ri)$ grows.
        }
\label{fig:WdW}
\end{figure}

Another application of~\eqref{v1} is an alternative boundary explanation of the volume growth of ``Wheeler-de Witt'' (WdW) patches of the eternal black hole.  Take $\sN_{\ft, w} = \sY_{L_{\ft, w}} \lor \sY_{R_{\ft, w}}$ to be the algebra associated with the union of time bands of width $w$ centered around 
time $\ft$ in the left and right CFTs about the thermofield double state $\ket{\Psi_\b}$ (above the Hawking-Page temperature). Take $\sM$ to be the full boundary algebra of the left and right CFTs in this sector. $\sN_{\ft, w}'\cap \sM$ is  dual to
the diamond  region $\bA_{\ft, w} = \hat \fa_{\ft, w}$ in  Fig.~\ref{fig:WdW}, with $\fa_{\ft, w}$ denoting the maximal volume slice. 
From~\eqref{v1}, the index $(\sM: \sN_{\ft, w})$ can be obtained from the volume ${\rm Vol} (\fa_{\ft, w})$. 
 
As we increase the time-band width $w$, ${\rm Vol} (\fa_{\ft, w})$ decreases. Now consider increasing $\ft$ from $0$, but with $w$ fixed. In particular, we can take $w$ to be infinitesimal, in which case, $\bA_{\ft, w}$ really becomes the WdW patch considered previously in the complexity=volume~\cite{Susskind:2014rva, Stanford:2014jda} {and complexity=action~\cite{Brown:2015bva} proposals}. The increase of ${\rm Vol} (\fa_{\ft, w})$ with $\ft$ was interpreted in that context as the increase of the complexity of the state $e^{- i (H_R + H_L) \ft} \ket{\Psi_\b}$ with time.
 
In terms of~\eqref{v1}, the volume increase with $\ft$ can now be interpreted as the growth of the relative commutant $\sN_{\ft, w}' \cap \sM$, or equivalently, the shrinking of the relative ``size'' of $\sN_{\ft, w}$ in $\sM$. Naively, one may have thought that since the size of the time-bands does not change with $\ft$, the algebra $\sN_{\ft, w}$ should remain the same. However, the embedding of $\sN_{\ft, w}$ into $\sM$ is changing with $\ft$ and therefore so does the index $(\sM: \sN_{\ft, w})$. In Sec.~\ref{sec:alg}, we give an explicit boundary derivation of the increase of $(\sM: \sN_{\ft, w})$ with $\ft$, which can be viewed as a self-consistency check of~\eqref{v1}. 

That the representation of the same algebra on the state $e^{- i (H_R + H_L) \ft} \ket{\Psi_\b}$
becomes smaller with increasing  $\ft$
may be viewed as a complementary perspective---in the Heisenberg picture---to the idea that the state 
$e^{- i (H_R + H_L) \ft} \ket{\Psi_\b}$
becomes more complex as $\ft$ grows.

\medskip

\noindent {\bf Plan of the paper}

\medskip

The plan of the paper is as follows. In section~\ref{sec:alg}, we demonstrate how the algebra of the union of finite left and right time-bands centered at the same boundary time, $\ft$, about the TFD state on two holographic CFTs ``shrinks'' as $\ft$ is increased. We argue that this algebra ``shrinking'' is dual to the volume growth of the WdW patch with increasing boundary anchor time. We then present a speculative connection between our algebra ``shrinking'' and the growth of complexity of the TFD state. Next, in section~\ref{sec:ex} we apply our proposal~\eqref{v1} to several examples, including finite bulk domains of dependence, {the region between the entanglement and casual wedges}, and cases of holographic superadditivity. In section~\ref{sec:finiteN}, we discuss possible interpretations of our index at a finite value of $N$. In section~\ref{sec:dS}, we consider algebras of observables for observers at the North and South poles of de Sitter (dS) space and discuss the ramifications of volume growth in dS for the relationships amongst observer algebras under the proposal~\eqref{v1}. Finally, we conclude with a discussion in section~\ref{sec:disc}. Details on the definitions of index of inclusion in general cases are collected in Appendix~\ref{app:index}, while useful results on half-sided modular translations and computations of volumes in AdS$_2$ are collected in Appendix~\ref{app:HSMT}.

\bigskip

\noindent \textbf{Notations and conventions:}

\smallskip

\begin{itemize}

\item $A'$ denotes the bulk/boundary causal complement of the bulk/boundary spacetime subregion $A$.

\item $\hat{A}$ denotes the bulk/boundary domain of dependence of a bulk/boundary region $A$. 

\item Unless mentioned otherwise, in this paper an algebra always means a von Neumann algebra of operators acting explicitly on a Hilbert space.

\item $\sN'$ denotes the commutant of the von Neumann algebra $\sN$.

\item $\sB (\sH)$ denotes the algebra of bounded operators on the Hilbert space $\mathcal{H}$.

\end{itemize}

\section{Algebraic description of volume growth of a black hole} \label{sec:alg}

In this section we first give a qualitative boundary explanation for the volume growth of a black hole. We then comment on a possible connection with the interpretation in terms of complexity growth.

\subsection{A boundary explanation of the volume increase} \label{sec:alG}

Consider the algebras  $\sN_{\ft, w}$ and $\sN_{0, w}$ associated with the union of left and right time bands of half width $w$ centered respectively around $t = \ft > 0$ and $t=0$ in the sector about the TFD state {at temperature above the Hawking-Page transition}.  In the formulation~\eqref{v1}, the increase in the bulk volume with time can be  interpreted as the increase in the size of the commutant of $\sN_{\ft, w}$ with $\ft$.  Below we argue that $\sN_{\ft, w}$ is unitarily equivalent to a proper subalgebra $\sN_{0, w_* (\ft)}$ of $\sN_{0, w}$, where $w_* (\ft)$  is a monotonically decreasing function of $\ft$ satisfying $w_* (\ft) < w,~ \forall \ft > 0$.

We first enumerate {a convenient family of} unitary transformations. 
In the large $N$ limit, the time-translations generated by the individual left and right CFTs Hamiltonians, $H_R$ and $H_L$, are not defined as $H_R$ and $H_L$ do not have a well-defined large $N$ limit. Their combination $K = H_R - H_L$ is, however, well-defined, and generates opposite time translations of the $R$ and $L$ CFTs.  
There are also ``hidden'' half-sided modular translations generated by operators $P_+,~P_- \in \sB(\sH_{\rm GNS})$ discussed in~\cite{longPaper}. 
These are emergent unitary transformations of the boundary system in the large $N$ limit. See Appendix~\ref{app:HSMT} for a collection of relevant formulas.  
Acting on boundary operators, these transformations are nonlocal. But in the limit large operator dimension, it becomes local for operators averaged over boundary spatial dimensions (i.e., operators with zero spatial momentum). 
More explicitly, with $U_\pm (s)= e^{i P_\pm s}$, we have (for all values of $s$ such that the argument of the corresponding logarithm is positive)~\cite{longPaper} 
\bega \label{t1}
U_\pm (s) \sO_{R} (t) U_\pm^\da (s) \propto \sO_R (f_R^\pm (t;s))  , \quad 
U_\pm (s) \sO_{L} (t) U_\pm^\da (s) \propto \sO_L (f_L^\pm (t;s))  ,  \\
f_R^+ (t;s) = t - \log\le(1-s e^{t}\ri) , \quad f_L^+ (t;s) = t + \log\le(1+s e^{-t}\ri), \\
f_R^- (t;s) = t + \log\le(1+s e^{-t}\ri) , \quad f_L^- (t;s) = t - \log\le(1- s e^{t}\ri) \ .
\label{t4}
\end{gather}

Using $U_+ (s_*)$ with 
\be
	s_* \equiv - \frac{\sinh \ft}{\cosh w} \ ,
\ee
followed by a boundary time-translation generated by $K$ (that translates $t_R$ and $t_L$ by the same amount in opposite directions), we can map 
$\sN_{\ft, w}$ to the algebra of a union of time bands on the $R$ and $L$ boundaries centered at $t=0$. However, such a transformation maps the time-bands on the $R$ and $L$ boundaries unevenly. The half-widths of the left and right time-bands after transformation by $U_+ (s_*)$ are
\bega 
	w_L (s_*)  = w + \frac{1}{2} \log\le(\frac{\cosh(\ft + w)}{\cosh(\ft - w)}\ri), \qquad w_R (s_*) = w - \frac{1}{2} \log\le(\frac{\cosh(\ft + w)}{\cosh(\ft - w)}\ri) \ , \\
w_L (s_*) = \le(1 + \tanh \ft\ri) w + O(w^2), \qquad w_R (s_*) = \le(1 - \tanh \ft\ri)w + O(w^2), \quad w \to 0^+ \ .	
\label{eq:evolvedWidths}
\end{gather} 
The left time-band has grown in width and the right has shrunk.  So it is not obvious how the algebra corresponding to the union of these two time bands compares with $\sN_{0, w}$.

The above discussion motivates using translations generated by  $P_+ + P_-$, which act on the $R$ and $L$ boundaries symmetrically. Since $P_+$ and $P_-$ do not commute, the action of the unitary transformation 
\be 
U(s) = e^{is (P_+ + P_-)}
\ee
on $\sO_R, \sO_L$ does not follow simply from~\eqref{t1}--\eqref{t4}. We leave the details to Appendix~\ref{app:HSMT} and only state the result here 
\be 
U(s) \sO_{L, R} (t) U^\da (s) \propto  \sO_{L, R} (f (t;s)), 
\quad 
f (t;s) =  t - \log\le[ \frac{\cos s - \sin s e^{t}}{\cos s + \sin s e^{-t}} \ri] \ .
\ee
Now choosing $s_*$ given by 
\be
s_* = - \ha \tan^{-1} \le({\sinh \ft \ov \cosh w} \ri)
\ee
we can unitarily transform $\sN_{\ft, w}$ to $\sN_{0, w_*(\ft)}$ with 
\bea
w_*(\ft) &= &  \log \le({\sqrt{\cosh^2 w + \sinh^2 \ft} + \sinh w \ov \cosh \ft }\ri) < w  \\
 & = & {w \ov \cosh \ft} + O (w^2) \quad (w \to 0^+) \ .
\eea
Under this evolution the time bands on both sides shrink. This shows that the algebra $\sN_{\ft, w}$ is unitarily equivalent to a proper subalgebra of $\sN_{0,w}$. Note that since we cannot currently calculate the index on the boundary directly, we cannot use the boundary theory to quantitatively ``predict'' how the bulk volume should increase with $\ft$. However, this argument shows that it should increase. 

In order to conclude that the containment $\sN_{0,w_*(\ft)} = U(s_*) \sN_{\ft, w} U(s_*)^{\da} \subset \sN_{0,w}$ gives a boundary dual of the growth of the WdW patch in the bulk, we must assume that the index~\eqref{v1} is unitarily invariant. This is because our proposal suggests that $\le(\sB(\sH_{\rm GNS}) : \sN_{\ft, w}\ri)$ encodes the volume of the WdW patch at time $\ft$ via~\eqref{v1}, while $\le(\sB(\sH_{\rm GNS}) : \sN_{0, w}\ri)$ encodes the `initial' volume. Unitary invariance of the index allows us to conclude that $\le(\sB(\sH_{\rm GNS}) : \sN_{\ft, w}\ri) = \le(U(s_*)\sB(\sH_{\rm GNS}) U(s_*)^{\da} : U(s_*)\sN_{\ft, w}U(s_*)^{\da}\ri) = \le(\sB(\sH_{\rm GNS}) : \sN_{0, w_*(\ft)}\ri)$ and since $\sN_{0,w_*(\ft)} \subset \sN_{0,w}$ we then have $\le(\sB(\sH_{\rm GNS}) : \sN_{\ft, w}\ri) \geq \le(\sB(\sH_{\rm GNS}) : \sN_{0, w}\ri)$, corresponding to the growth of the volume of the WdW patch with $\ft$. While we cannot explicitly demonstrate boundary unitary invariance for our index, it can be explicitly shown for the Kosaki index: i.e. one can show that $[U\sM U^{\da} : U\sN U^{\da}] = [\sM : \sN]$ for any unitary $U \in \sB(\sH)$.
As a consistency check on our index we should find that the volume of the maximal slice of the dual to $\sN'_{\ft,w} \cap \sM$ agrees with that 
of $\sN'_{0,w_*(\ft)} \cap \sM$ for the case of AdS$_2$.\footnote{This explicit check can only be done for AdS$_2$ as this is the only case in which we know the explicit form of the unitary evolution for {\bf all} operators in $\sN_{\ft, w}$ at large scaling dimension. In other dimensions, only transformations on zero spatial momentum operators are known.} This can indeed be checked explicitly. 
In this case, the half-sided modular translations $P_+, P_-$ generate isometries, and thus do indeed preserve the volume of the dual. See Appendix~\ref{app:HSMTb} for details.

\subsection{Possible connection to complexity of the state}

We now discuss a possible connection between our index and the complexity of the state. Applying the complexity=volume proposal~\cite{Susskind:2014rva, Stanford:2014jda} {to the eternal black hole setting,}
the bulk volume growth is identified with the increase of complexity of the state 
\be \label{timE}
  \ket{\Psi_\b (\ft)} =  e^{-i (H_R + H_L) \ft} \ket{\Psi_\b}  \ .
\ee

An important point to keep in mind is that, in the large $N$ limit, the states $\ket{\Psi_\b (\ft)}$ and $\ket{\Psi_\b} \equiv \ket{\Psi_\b (0)}$ do not lie in the same Hilbert space, since the operator $ e^{-i (H_R + H_L) \ft}$ does not have a well-defined large $N$ limit for any nonzero $\ft$.  
More explicitly, denote the full algebra of single-trace operators of $R$ and $L$ CFTs as $\sS$, which can be viewed as an abstract $C^*$-algebra with norm inherited from the theory at finite $N$. 
Acting $\sS$ on $\ket{\Psi_\b}$ 
and $\ket{\Psi_\b (\ft)}$ leads to Hilbert spaces of ``low energy'' excitations (called GNS Hilbert spaces), which we denote respectively as $\sH$ and $\sH_\ft$. 
$\sH$ and $\sH_\ft$ do not overlap, as the time evolution in~\eqref{timE} cannot be obtained by the action of single-trace operators.  
Denote the representations of $\sS$ in these Hilbert spaces respectively as $\pi$ and $\pi_\ft$. For example, consider an operator $\sO \in \sS$, then $\pi (\sO)$ denotes its representation on $\sH$ and $\pi_t (\sO)$ on $\sH_{\ft}$. 

Now consider the following time-band algebras acting on $\sH_{\ft_1}$ built around the state $\ket{\Psi_\b (\ft_1)}$, 
\be
\sN_{\ft_2, w}^{(\ft_1)} 
= \pi_{\ft_1} (\sS_{L_{\ft_2, w}})  \lor \pi_{\ft_1} (\sS_{R_{\ft_2, w}}) \ ,
\ee
where $\sS_{L_{\ft_2,w}}$ is algebra of single-trace operators for the time-band $L_{\ft_2, w}$ centered at $t= \ft_2$ with width $w$ on the left boundary and similarly for $\sS_{R_{\ft_2,w}}$ on the right. We also denote as $\sN_{\ft_2}^{(\ft_1)}$ the representation of the algebra for the union of left and right half-infinite time-bands starting at $\ft_2$ 
acting on $\sH_{\ft_1}$.

While {$\pi (\sS)$ and $\pi_\ft (\sS)$}  act on different Hilbert spaces, there is an algebraic isomorphism\footnote{Namely, a bijective map that preserves products, linear combinations, and the adjoint operation.} between them. Note the relation 
\be
\vev{\Psi_\b|\sO_1 (t_1) \sO_2 (t_2)|\Psi_\b}  = \vev{\Psi_\b (\ft)|\sO_1 (t_1-\ft) \sO_2 (t_2-\ft)|\Psi_\b (\ft)} , \quad \sO_1, \sO_2 \in \sS \ ,
\ee
where $\sO_1, \sO_2$ can be arbitrary products of single-trace operators (regardless of whether they are from $R$ or $L$ CFTs). 
Since the inner products and the definitions of $\pi (\sO)$ and $\pi_\ft (\sO)$ in $\sH$ and $\sH_\ft$ only depend on the 
two-point functions of the single-trace operators in the corresponding state, there is a bijective mapping $b_{0,\ft} : \pi_0 (\sS) \equiv \pi_{\ft = 0}(\sS) \to \pi_{\ft}(\sS)$ given by
\be
 b_{0,\ft} \le(\pi_0 (\sO(t))\ri) = \pi_\ft (\sO (t-\ft)), \quad b_{\ft,0} \le(\pi_{\ft} (\sO(t))\ri) = \pi_0 (\sO (t+\ft)) \ ,
\ee
that preserves the algebraic relations of these operators in their respective Hilbert spaces.
We then have the bijective map $b_{\ft_1, \ft_1 + \ft} : \pi_{\ft_1}(\sS) \to \pi_{\ft_1 + \ft}(\sS)$ leading to the isomorphisms 
\bega \label{map1}
b_{\ft_1, \ft_1 + \ft} \le(\sN_{\ft_2, w}^{(\ft_1)} \ri) = \sN_{\ft_2-\ft, w}^{(\ft_1 + \ft)} , \quad b_{\ft_1 + \ft, \ft_1 } \le(\sN_{\ft_2 - \ft, w}^{(\ft_1 + \ft)} \ri) = \sN_{\ft_2, w}^{(\ft_1)} \\
b_{\ft_1, \ft_1 + \ft} \le( \sN_{\ft_2}^{(\ft_1)} \ri) = \sN_{\ft_2-\ft}^{(\ft_1 + \ft)}, \quad b_{\ft_1 + \ft, \ft_1} \le( \sN_{\ft_2 - \ft}^{(\ft_1 + \ft)} \ri) = \sN_{\ft_2}^{(\ft_1)}  \ .
\label{map2}
\end{gather}

$\sN_{0,w}^{(\ft)}$ gives the representation of the two-sided time-band algebra centered at $t =0$ with width $w$ in the Hilbert space of $\ket{\Psi_\b (\ft)}$. As we vary $\ft$, it gives a family of representations of the {\it same} algebra.  Similarly for $\sN_0^{(\ft)}$. 
 Now from~\eqref{map2}, we have 
\bega 
b_{\ft, 0} \le(\sN_{0}^{(\ft)}\ri) = \sN_{\ft}^{(0)} \subset \sN_{0}^{(0)} , \quad \ft > 0, \\
b_{\ft_1, 0} \le(\sN_{0}^{(\ft_1)}\ri) = \sN_{\ft_1}^{(0)} \subset \sN_{\ft_2}^{(0)} = b_{\ft_2, 0} \le( \sN_{0}^{(\ft_2)} \ri) , \quad \ft_2 < \ft_1\ .
\end{gather}
This shows how there is a sense in which the representation of the algebra of the union of left and right semi-infinite time-bands $t \in (0, \infty)$ around the state $\ket{\Psi_\b (\ft)}$ `shrinks' as $\ft$ is increased.
Similarly, from~\eqref{map1} and the discussion of Sec.~\ref{sec:alG}, we have\footnote{Here we use the notation ``$\simeq$'' to denote unitary equivalence, which is a stronger equivalence than algebraic isomorphism.}
\bega
b_{\ft, 0} \le(\sN_{0, w}^{(\ft)} \ri) = \sN_{\ft, w}^{(0)}  \simeq  \sN_{0, w_*(\ft)}^{(0)}  \subset  \sN_{0, w}^{(0)}   ,  \\
b_{\ft_1, 0} \le(\sN_{0,w}^{(\ft_1)} \ri) =  \sN_{\ft_1,w}^{(0)} \simeq  \sN_{0, w_* (\ft_1)}^{(0)} \subset 
\sN_{0, w_* (\ft_2)}^{(0)} \simeq 
\sN_{\ft_2,w}^{(0)} = b_{\ft_2, 0} \le( \sN_{0,w}^{(\ft_2)} \ri), \quad \ft_2 < \ft_1\ .
\end{gather} 
We thus see that as we increase $\ft$, there is a sense in which  $\sN_{0,w}^{(\ft)}$ shrinks. 
That the representation of the same algebra on the state $\ket{\Psi_\b (\ft)}$ becomes smaller with $\ft$ may be  
seen as a reflection that 
the state $\ket{\Psi_\b (\ft)}$ becomes more complex with $\ft$. 
This provides a complementary perspective on the growth of complexity, in the Heisenberg picture.

\section{Other applications} \label{sec:ex}

In this section we discuss a number of other applications of~\eqref{v1}. {We demonstrate how the proposal~\eqref{v1} helps us to quantitatively characterize relationships between algebras in several cases of interest.} 

\subsection{Interpretation of the volume of a bulk diamond} 

\begin{figure}[h]
        \centering
		\includegraphics[width=0.30\textwidth]{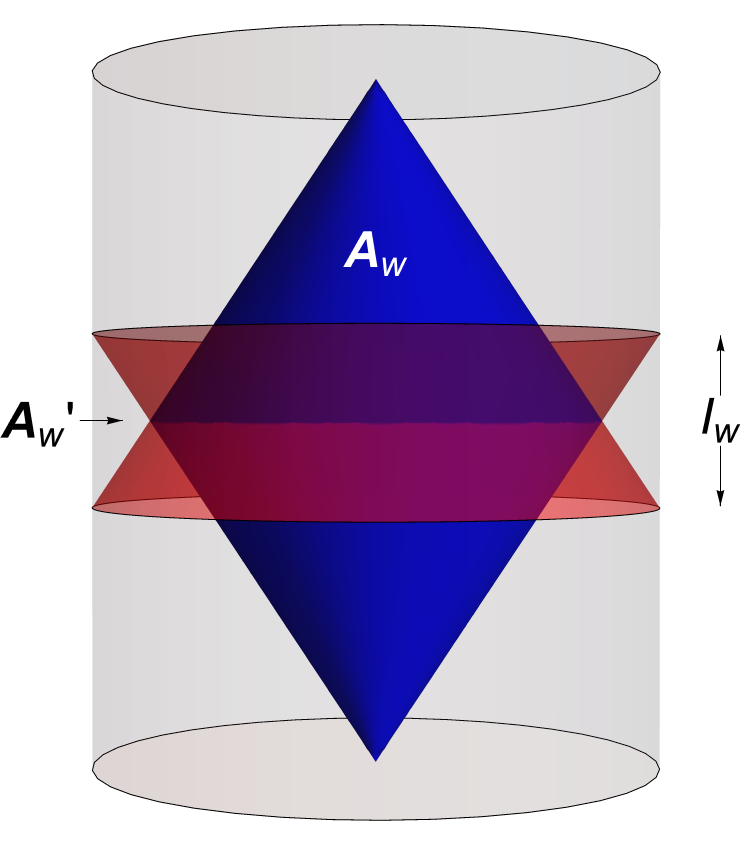}
		\quad
		\includegraphics[width=0.45\textwidth]{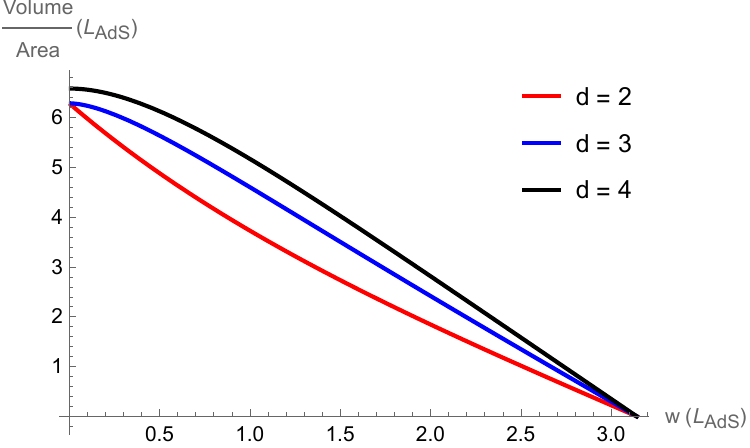}
                \caption[  ]
        {\small Left: Bulk dual, $\bA_w',$ of the algebra, $\sN$, of the time-band $I_w$ (shaded gray wedge region bounded by red surfaces) and bulk dual, $\bA_w$, of the relative commutant $\sN' \cap \sM$ (blue region). Right: Ratio of volume of $\fa_w$ to area of $\pt \fa_w$ in units of $l_{\rm AdS}$.
        }
\label{fig:diamond}
\end{figure}

Consider the example of Fig.~\ref{fig:diamond}(a), where we take $\sN$ to be the boundary algebra of a time-band $I_w$ of width $w$ and $\sM$ to be the full boundary algebra in the vacuum sector. 
$\sN$ is dual to the bulk wedge region $\bA_w'$, and the relative commutant 
$\sN' \cap \sM$ is dual to the bulk diamond region $\bA_w = \hat \fa_w$, which has maximal volume slice $\fa_w$. 
 As we increase $w$, $\sN$ grows and $(\sM: \sN)$ should decrease. {We explicitly see this through} the decrease of ${\rm Vol}(\fa_w)$ with the shrinking of $\fa_w$. As $w$ approaches $\pi l_{\rm AdS}$,\footnote{The boundary sphere is taken to have radius $l_{\rm AdS}$.} where $l_{\rm AdS}$ is the radius of AdS, $\fa_w$ shrinks to zero size. More explicitly, 
the volume of $\fa_w$ is
\be 
	\vol(\fa_w) = {l_{\rm AdS}^d \vol(S^{d-1}) \ov d} (\sin^2 \rho_w)^{{d \ov 2}} F\le({d \ov 2}, {d+1 \ov 2} ; {d\ov 2} + 1 ; \sin^2 \rho_w \ri) , 
\label{eq:VolHolePureAdS}
\ee
where $\rho_w$ is related to the ``proper radius'', $r_w,$ of $\fa_w$, by
\be \label{timebad}
 r_w = l_{\rm AdS} \tan\le(\rho_w\ri), \qquad \rho_w = \frac{\pi l_{\rm AdS} - w}{2l_{\rm AdS}}
  \ .
 \ee
{We emphasize that the decrease in $(\sM: \sN)$ as $w$ is increased is completely distinct from the decrease in entanglement entropy between $\sN$ and $\sN'$.}
In contrast {to the volume}, the area of $\pt\fa_w$ is given by
\be 
{\rm Area} (\pt\fa_w) = l_{\rm AdS}^{d-1} \tan^{d-1} \rho_w \ .
\ee
The area is believed to capture the entanglement between $\sN$ and $\sN'$, i.e., between the diamond $\hat \fa_w$ and its complement. We plot the ratio of volume to area (in units of $l_{\rm AdS}$) in Fig.~\ref{fig:diamond}(b). This ratio goes linear for large enough $w$ since we are then comparing volume to area of balls in AdS that are smaller than the AdS scale so the result for the ratio agrees to that in flat space. {This ratio is not constant, so we see that $(\sM: \sN)$ is changing in a distinct manner from the entanglement between $\sN$ and $\sN'$.}

\begin{figure}[h]
        \centering
		\includegraphics[width=5cm]{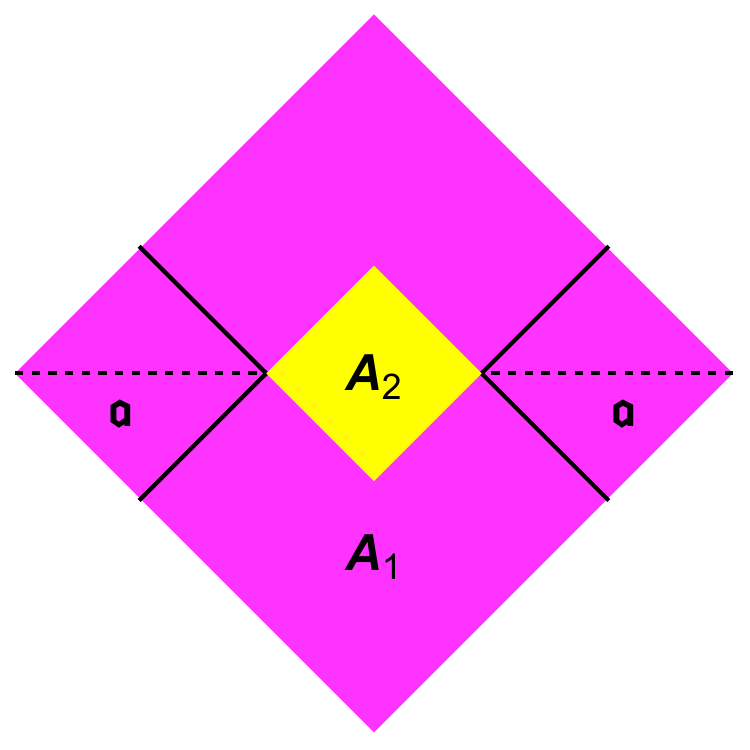} \quad
		\includegraphics[width=5cm]{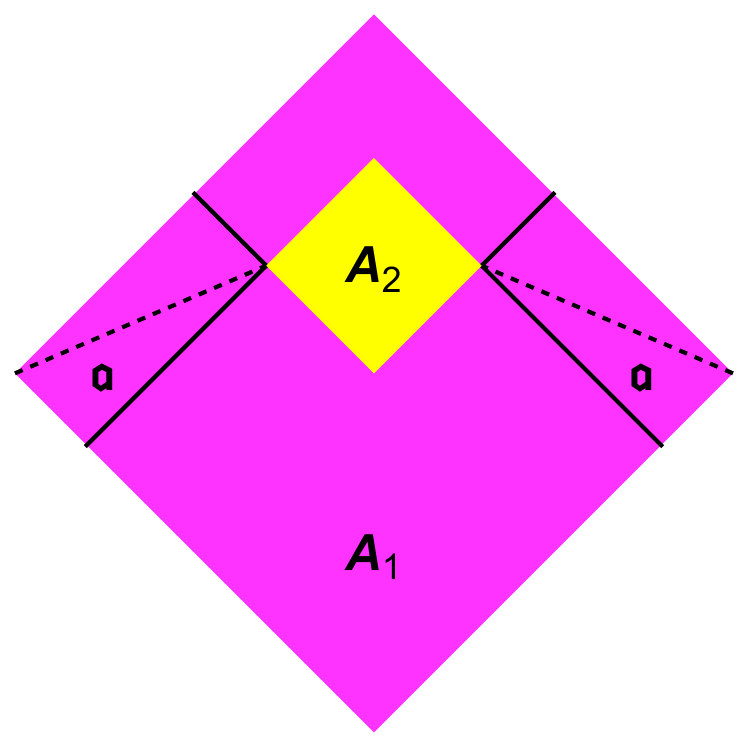}
                \caption[  ]
        {\small Left: Nested bulk diamonds $\bA_2 \subset \bA_1$ whose maximal volume slices lie on the same time-reflection symmetric bulk time slice. Right: Nested bulk diamonds $\bA_2 \subset \bA_1$ whose maximal volume slices lie on different time-reflection symmetric bulk time slices. Even though ${\rm Vol}(\fa_{1,2})$ are each the same in both the left and right figures, the index is very different in the two cases.
        }
\label{fig:embed}
\end{figure}

{Another important aspect of the index is that it depends on the particular embedding of the subalgebra into the larger algebra}. Now consider the algebras $\sN_2 \subset \sN_1$ corresponding to nested bulk diamonds $\bA_2 \subset \bA_1$ in empty AdS$_{d+1}$. The index $(\sN_1: \sN_2)$ is then captured by the volume of the maximal volume slice $\fa$ of $\bA_2' \cap  \bA_1$. As indicated in Fig.~\ref{fig:embed}, depending on how $\bA_2$ is embedded in $\bA_1$, or equivalently how $\sN_2$ is embedded in $\sN_1$, the index can be very different.

\subsection{Quantifying relative size of entanglement wedge and causal wedge algebras}

\begin{figure}[h]
        \centering
        \includegraphics[width=6cm]{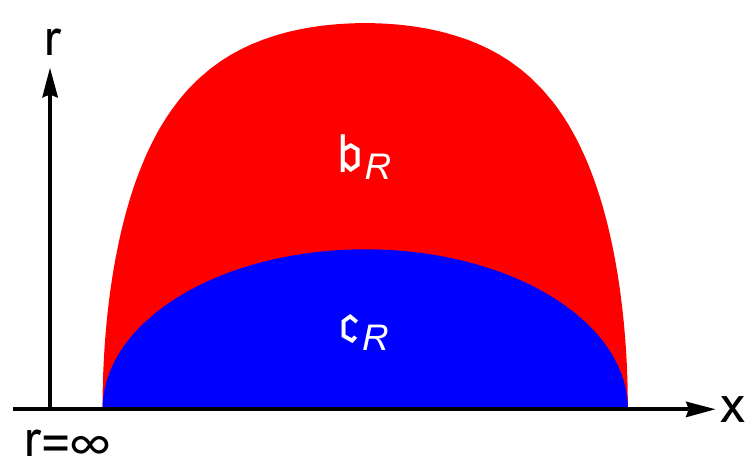}
                \caption[  ]
        {\small A time-reflection symmetric bulk Cauchy slice. The maximal volume slices, $\fb_R$ (union of red and blue) and $\fc_R$ (blue), respectively, of the entanglement and causal wedges for a spatial boundary subregion $R$ lying on a time-reflection symmetric slice. 
        }
\label{fig:entcau}
\end{figure}

Consider a boundary spatial subregion $R$. Two different subalgebras can  be naturally assigned to $R$ in the large $N$ limit: the algebra $\sX_R$ of all operators in $R$ that survive the large $N$ limit, and the algebra $\sY_{\hat R} \subseteq \sX_R$, generated  by single-trace operators in $\hat R$. $\sX_R$ and $\sY_{\hat R}$ are respectively dual to the entanglement wedge $\hat \fb_R$ and causal wedge\footnote{We define causal wedge to be the causal completion of the standard definition.} $\hat \fc_R$ of $R$. 
From~\eqref{v1}, the index $(\sX_R : \sY_{\hat R})$ is then given by the volume of the maximal volume slice of $(\hat \fc_R)' \cap \hat \fb_R$. For simplicity, we will take $R$ to lie on a time-reflection symmetric slice, in which case $\fb_R$ and $\fc_R$ lie on the same bulk time-reflection symmetric slice, and the maximal volume slice of $(\hat \fc_R)' \cap \hat \fb_R$ is simply given by $\overline{\fc_R} \cap \fb_R$.\footnote{Here, $\overline{\fc_R}$ denotes the complement of $\fc_R$ within the time-reflection symmetric bulk Cauchy slice.} See Fig.~\ref{fig:entcau}.

It can be shown that in general the volume of $\overline{\fc_R} \cap \fb_R$, i.e., ${\rm Vol} (\overline{\fc_R} \cap \fb_R) =  {\rm Vol} (\fb_R) 
- {\rm Vol} (\fc_R ) $ is infinite for $d \geq 3$. With $z= \ep$ ($0 < \ep << 1$) being the AdS boundary cutoff, both volumes can have bulk IR divergences of the form $\#\ep^{1-d} + \#\ep^{3-d} + \cdots$, but in general only the leading divergence agrees.\footnote{For odd $d$ the smallest non-vanishing term as $\ep \to 0$ is a logarithmic divergence, while for even $d$ it is a constant.} Thus generically ${\rm Vol} (\fb_R) - {\rm Vol} (\fc_R ) $ diverges as $\ep^{3-d}$ for $d \geq 4$ and as $\log(\ep)$ for $d=3$.  This is because the boundaries of the entanglement and causal wedges agree only up to $O(z)$ in the bulk. From the boundary perspective this 
implies that in $d \geq 3$ there are a large number of extra operators {in the entanglement wedge algebra} at short-distances such that the index is UV sensitive. 
For $d=2$, the volume differences are finite. Here  we mention three explicit results:

\subsubsection{Two disjoint intervals in vacuum} \label{ex:two}

Consider the union of two boundary intervals $R \equiv R_1 \cup R_2$ on a time-slice of a CFT$_2$ on $\RR \times S^1$ in the vacuum state, where $R_1 = \{(t,\theta) ~|~ t=0,~ -\fw < \theta + \frac{\pi}{2} < \fw\}$ and $R_2 = \{(t,\theta) ~|~ t=0,~ -\fw < \theta - \frac{\pi}{2} < \fw\}$ for $0 < \fw < \frac{\pi}{2}$. The entanglement wedge exceeds the causal wedge for $\fw > \frac{\pi}{4}$. The volumes of $\fb_R$ and $\fc_R$ were computed in~\cite{Ben-Ami:2016qex, Abt:2017pmf} and found to be $4\fw {l_{\rm AdS}^2 \ov \ep}$ and $4\fw {l_{\rm AdS}^2 \ov \ep} - 2\pi l_{\rm AdS}^2$, respectively, which gives
\be 
	\vol\le( \overline{\fc_R} \cap \fb_R\ri) = 2\pi l_{\rm AdS}^2 \ .
\ee
Note that the difference is {finite and} independent of the size of boundary circle and $\fw$.

\subsubsection{Single interval at finite temperature}

Consider a boundary interval $R=\{(t,\theta) ~|~ t=0,~ -\fw < \theta < \fw\}$ for a CFT$_2$ on $\mathbb{R} \times S^1$ at finite temperature above the Hawking-Page transition. The bulk dual is a BTZ black hole. The entanglement wedge of $R$ exceeds its causal wedge~\cite{Ryu:2006bv,Hubeny:2013gta} for $\fw > \fw_c = \frac{\beta}{2\pi l} \coth^{-1}\le(2\coth \le(\frac{2\pi^2 l}{\beta}\ri) - 1\ri)$ with $l$ the radius of the boundary circle and $\beta$ the inverse temperature of the boundary system.
The volumes of $\fb_R$ and $\fc_R$ are found to be $2\fw {l_{\rm AdS}^2 \ov \ep} + \pi l_{\rm AdS}^2$ and $2\fw {l_{\rm AdS}^2 \ov \ep} - \pi l_{\rm AdS}^2$~\cite{Ben-Ami:2016qex, Abt:2017pmf}, respectively, which gives
\be 
	\vol\le( \overline{\fc_R} \cap \fb_R\ri) = 2\pi l_{\rm AdS}^2 \ .
\ee
It is curious that this is independent of $l$, $\b$, and $\fw$. 

\subsubsection{Strip in $d \geq 3$}

Consider a strip of half-width $a$ on the $t=0$ slice of $d$-dimensional Minkowski space, $R=\{(t,x^i) ~|~ t=0,~ |x^1| < a,~ x^i \in \mathbb{R},~ i \in [2,3,...,d-1]\}$. About the vacuum state, the maximal volume slices of the entanglement and causal wedges in Poincar\'e AdS$_{d+1}$, $\fb_R$ and $\fc_R$ respectively, are computed in~\cite{Hubeny:2012wa}. Explicitly computing the volumes for this special case, one finds that $\vol(\fb_R)$ only contains the leading $\vol(R)\ep^{1-d}$ bulk IR divergence, while $\vol(\fc_R)$ has finite coefficients for all subleading bulk IR divergences. One therefore has that $\vol(\overline{\fc_R} \cap \fb_R)$ has leading bulk IR divergence of $\ep^{3-d}$ (or $\log(\ep)$ for $d=3$). Our index $\le(\sX_R : \sY_{\hat R}\ri)$ for the containment of the causal wedge algebra in the entanglement wedge algebra will therefore be divergent in this case.

\subsection{Quantifying superadditivity of boundary subalgebras} 

In~\cite{addPaper} the violation of additivity of boundary algebras in the large $N$ limit was discussed in detail.
For example, with $R_1$ and $R_2$ being subregions of the same boundary Cauchy slice, we can have $\sX_{R_1} \vee \sX_{R_2} \subsetneq \sX_{R_1 \cup R_2}$.
We can quantify how much larger $\sX_{R_1 \cup R_2}$ is than $\sX_{R_1} \vee \sX_{R_2}$ by computing the volume of the region between their entanglement wedges, $\overline{(\fb_{R_1} \cup \fb_{R_2})} \cap \fb_{R_1 \cup R_2}$. See Fig.~\ref{fig:superadd}. Here are some explicit examples:

\begin{figure}[h]
        \centering
		\includegraphics[width=0.4\textwidth]{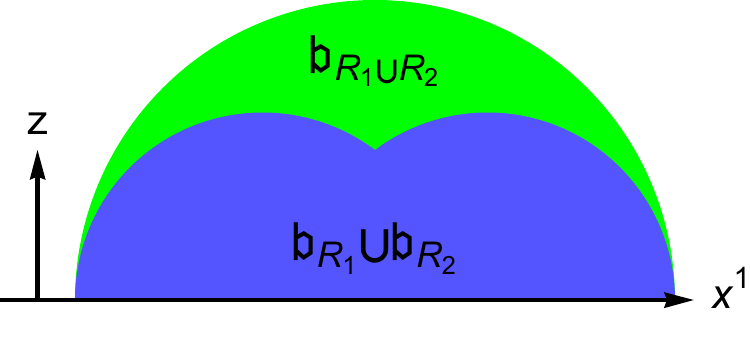} \hfill
		\includegraphics[width=0.47\textwidth]{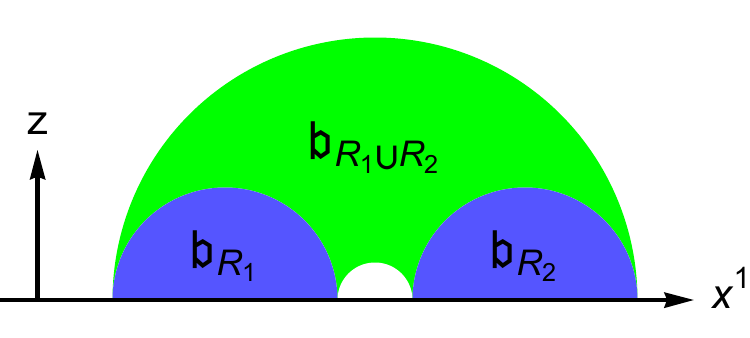}
                \caption[  ]
        {\small Quantifying superadditivity, $\sX_{R_1} \vee \sX_{R_2} \subsetneq \sX_{R_1 \cup R_2}$, using bulk volumes. A time-slice of AdS$_3$ is shown with the maximal volume slice, $\fb_{R_1} \cup \fb_{R_2}$, of the dual of $\sX_{R_1} \vee \sX_{R_2}$ shown in blue and that, $\fb_{R_1 \cup R_2}$, of $\sX_{R_1 \cup R_2}$ as the union of blue and green regions. Left: Overlapping boundary intervals, $R_1 \cap R_2 \neq \emptyset$. Right: Disjoint boundary intervals, $R_1 \cap R_2 = \emptyset$.
        }
\label{fig:superadd}
\end{figure}

\subsubsection{Two intersecting intervals for vacuum CFT$_2$}

Take two intervals of lengths $2a$ with centers separated by distance $2b$, with $a>b$. Explicitly, we have 
$R = \{(t,x) ~|~ t=0,~ x \in (-a-b,a+b)\}$, $R_1 = \{(t,x) ~|~ t=0,~ x \in (-a-b,a-b)\}$, $R_2 = \{(t,x) ~|~ t=0,~ x \in (-a+b,a+b)\}$. Their respective entanglement wedges are shown in the left panel of Fig.~\ref{fig:superadd}. The regulated volumes of these slices are 
\bega \label{eq:PCAdSVolInt}
	\vol\le(\fb_{R_{1,2}}\ri) = \frac{2a l_{\rm AdS}^2}{\ep} - \pi l_{\rm AdS}^2, \qquad \vol\le(\fb_{R}\ri) = \frac{2(a+b) l_{\rm AdS}^2}{\ep} - \pi l_{\rm AdS}^2 , \\
\vol\le( \fb_{R_1} \cup \fb_{R_2} \ri) = \frac{2(a+b) l_{\rm AdS}^2}{\ep} - \pi l_{\rm AdS}^2 - 2l_{\rm AdS}^2 \arctan\le( \frac{b}{\sqrt{a^2-b^2}} \ri) ,
\label{eq:PCAdSVolAdd}
\end{gather} 
which gives 
\be 
	\vol\le( \overline{(\fb_{R_1} \cup \fb_{R_2})} \cap \fb_R \ri) = 2l_{\rm AdS}^2 \arctan\le( \frac{b}{\sqrt{a^2-b^2}} \ri)  \ .
\label{eq:PCAdSVolRCIntDecomp}
\ee
Notice that~\eqref{eq:PCAdSVolRCIntDecomp} vanishes as $b \to 0^+$. This can be understood since $R_1$ and $R_2$ become identical in this limit, i.e. $R_{1,2} \to R$, and there is no longer any superadditivity. Alternatively, as $b \to a^-$, $R_1$ and $R_2$ are almost separating and will overlap only at a single boundary point. In this limit~\eqref{eq:PCAdSVolRCIntDecomp} goes to $\pi l_{\rm AdS}^2$ as superadditivity persists.

\subsubsection{Two disjoint intervals for vacuum CFT$_2$}

Now consider the same configuration as above with $b > a$, for which the two intervals are disjoint. 
For $a < b < \sqrt{2} a$, $\fb_{R_1 \cup R_2}$ is simply connected and this story now reduces to the example in item~\ref{ex:two}, giving
\be 
	\vol\le( \overline{(\fb_{R_1} \cup \fb_{R_2})} \cap \fb_R \ri) = 2\pi l_{\rm AdS}^2 \ .
\label{eq:PCAdSVolRCTwoIntConn}
\ee
Notice that~\eqref{eq:PCAdSVolRCTwoIntConn} for $b \to a^+$ is twice the limit of~\eqref{eq:PCAdSVolRCIntDecomp} for $b \to a^-$, indicating that the measure of superadditivity is discontinuous as a function of the separation between the centers of $R_1$ and $R_2$ and is thus sensitive to the topology of $R_1 \cup R_2$.
For $b > \sqrt{2} a$ there is no longer superadditivity as we have $\sX_{R_1} \vee \sX_{R_2} = \sX_R$.

\section{Possible finite $N$ interpretation of the index} \label{sec:finiteN}

\subsection{Further elaboration on the index formula} 

In equation~\eqref{v1}, we proposed that the bulk volume can be used to obtain an index for boundary algebra inclusions. However, there are a number of issues that should be further clarified.

There are two ways of interpreting equation~\eqref{v1}: 

\ben 

\item It applies to algebras that are defined only in the large $N$ limit. 

\item It applies to algebras that can be intrinsically defined at finite $N$, and gives the leading order behavior of the indices of such algebras in the large $N$ limit.  This would be analogous to the black hole entropy formula. 

\een

First consider interpretation 1 above, which appears natural as all the algebras in our earlier discussion have been 
formulated in the large $N$ limit. With these algebras being type III$_1$, we can in principle define the Kosaki (or equivalently Longo) index for their inclusions. However, for the inclusions of under discussion, it is expected that the Kosaki indices are infinite
in the continuum limit.\footnote{
In the continuum limit of the bulk EFT the algebra inclusions we are describing are split inclusions of type III$_1$ factors, for which the normal conditional expectation needed for a finite Kosaki index does not exist. See, for example, section 6 of~\cite{Faulkner:2024gst} for more details.
} 
To have a finite index, we can consider introducing a {\it bulk cutoff},  $\ep$, to regularize 
the algebras to type I, and accordingly the constant $C$ in the exponent of~\eqref{v1} should have  the form 
\be \label{cu1}
C = {c \ov \ep^a l_{\rm AdS}^b}
\ee
where $c$ is some dimensionless constant, and $a+b = d$. At this stage it is not completely clear what $a$ should be. 
One natural candidate is $a=d,~b=0$, corresponding to a uniform lattice in all bulk directions. Another, holographically motivated, candidate is $a =d-1,~ b =1$, where the discretization of the bulk radial direction is related to the boundary time-resolution.

Now consider interpretation 2, which requires the existence of algebras that can be defined at finite $N$. So far, most algebras discussed in Sec.~\ref{sec:alg} and Sec.~\ref{sec:ex}---such as time-band algebras and causal wedge algebras---have only been defined precisely in the $N \to \infty$ limit. Below we present an argument based on the generalized gravitational entropy that these algebras may also be defined at finite $N$ and become type I.  If they can indeed be defined at finite $N$, then it is natural for the constant $C$ in~\eqref{v1} to have the form 
\be \label{cu2} 
C = {c \ov G_N  l_{\rm AdS}}
\ee
where $c$ is again some dimensionless number.  Equation~\eqref{cu2} may be also interpreted from~\eqref{cu1} with $a =d-1$ and $\ep = \ell_{p}$, the Planck length.

Below we present arguments for the existence of the algebras at finite $N$ and~\eqref{cu1} with $a=d-1$. The argument is somewhat speculative, but nevertheless suggestive.

\subsection{Algebras at finite $N$ and generalized gravitational entropy}

We now argue that the algebras discussed in Sec.~\ref{sec:ex} can be extended to finite $N$. 
We will use the time-band algebra $\sY_I$ associated with a boundary time-band $I$ in the vacuum sector as an illustration. 
In the large $N$ limit, $\sY_I$ is dual to the bulk spherical Rindler region (see the red region of Fig.~\ref{fig:diamond} Left) $\bB = \hat \fb$, with 
the identification $\wt \sM_\fb = \sY_I$~($\wt \sM_\fb$ is the bulk algebra in $\fb$). 
 Since $I$ covers a full Cauchy slice on the boundary, a possible finite-$N$ extension of $\sY_I$ is $\sB (\sH_{\rm CFT})$, the full boundary operator algebra. We now argue that there could exist a finite $N$ extension of $\sY_I$ that 
 is smaller than $\sB (\sH_{\rm CFT})$.

For this purpose, consider first the bulk algebra $\wt \sM_\fb$  in the $G_N \to 0$ limit, i.e. perturbatively in $G_N$. 
It is type III$_1$ in the continuum, and an entropy cannot be defined. 
Nevertheless, it is generally believed (see for example appendix A of~\cite{Bousso:2015mna} for a review) that the generalized gravitational entropy 
\be \label{gen2} 
S_{\rm gen} [\fb] = \lim_{\ep \to 0} \left( {{\rm Area} (\p \fb) \over 4 G_N (\ep)} + \wt S_{\fb} (\ep) \right) 
\ee
exists. Here $\ep$ is a bulk short-distance cutoff, $\wt S_{\fb} (\ep)$ is the bulk entanglement entropy for $\fb$, and $G_N (\ep)$ is the bare Newton constant (viewed as a coupling in the bulk effective field theory). 

In the presence of a bulk cutoff $\ep$, the bulk algebra $\wt \sM_\fb$ turns into a type I algebra $\wt \sM_\fb^\ep$, which now has an entropy $\wt S_{\fb} (\ep)$. The entropy $\wt S_\fb (\ep)$ is divergent in the $\ep \to 0$ limit, with the leading divergence of the form 
\be \label{div1}
\wt S_\fb (\ep) = f  {{\rm Area} (\p \fb) \over \ep^{d-1}} + \cdots, 
\ee
where $f$ is some constant and $\cdots$ denotes less divergent and finite terms in the limit $\ep \to 0$. 
The specific form of the cutoff, e.g., a lattice regularization, is not important. The bare coupling $G_N (\ep)$ is also divergent in the $\ep \to 0$ limit. 
Equation~\eqref{gen2} states that the total entropy is well defined despite the fact that each term on the right hand side cannot be individually defined in the continuum limit.

From subregion-subalgebra duality, we should have 
\be \label{sub1}
\wt \sM_\fb^\ep = \sY_I^\ep .
\ee
Namely, there exists a regularization on the boundary in the $N \to \infty$ limit that turns $\sY_I$ into a type I algebra $\sY_I^\ep$. While  we do not currently know how to describe the regularization explicitly for a general boundary theory, the duality implies that it must exist.  Note that at this stage $\ep$ is independent of {the renormalized value of} $G_N$.

The existence of a well-defined limit for equation~\eqref{gen2} can be interpreted as the existence of an extension of generalized gravitational entropy $S_{\rm gen} [\fb]$ to finite $G_N$, just as in the case of black hole entropy. Here by finite $G_N$ (or $N$), we always mean treating $G_N$ (or $1/N$) non-perturbatively. It is then natural to interpret $S_{\rm gen} [\fb]$ as the entropy for a finite $N$ extension of 
the time-band algebra $\sY_I^\ep$. After all, when $\ep \sim \ell_p$, one can no longer separate the area term in~\eqref{gen2}
and the term~\eqref{div1} in $\wt S_\fb (\ep)$. We denote this {sector-dependent} finite-$N$ extension as $\sB_I$. $\sB_I$ cannot be the full boundary algebra $\sB (\sH_{\rm CFT})$ as  $\sB (\sH_{\rm CFT})$ has zero entropy in the vacuum state. 
 Note that at finite $N$, the region $\fb$ can no longer be sharply defined due to spacetime fluctuations. 
What we are saying is that there is nevertheless a precisely-defined algebra. 
We stress that the definition of $\sB_I$ is sector-dependent, as the physics associated with the bulk dual region $\fb$ is different 
in empty AdS and in a black hole geometry (or any bulk dual with a different geometry in $\fb$).

\section{Algebraic probe of de Sitter volume growth} \label{sec:dS}
We now use our proposal~\eqref{v1} to describe a probe of the volume growth of de Sitter space formulated in terms of restricted algebras of observables associated to observers fixed at the North and South poles. We first review the static and Kruskal coordinate systems on de Sitter (dS) space. We then describe the relevant algebras of observables associated to observers travelling along the worldlines of the North and South poles. Finally, we apply our proposal and discuss how the growth of the volume of the region between the North and South cosmological horizons implies a particular behavior of the observers' algebras.

\subsection{Kruskal and Static Coordinates on dS$_{d+1}$}
\begin{figure}[h]
        \centering
        \includegraphics[width=6cm]{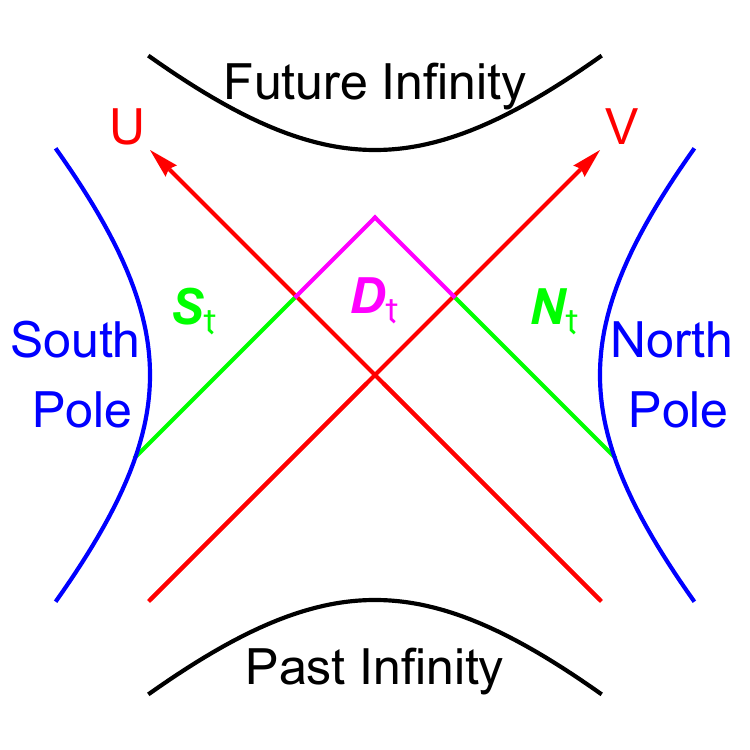}
                \caption[  ]
        {\small {The ``Kruskal'' diagram of de Sitter space. The cosmological horizons for the North and South pole observers are described in Kruskal coordinates by $U=0$ or $V=0$ and are depicted in red in the figure. The regions accessible to a North or South pole observer after proper time $\ft$ are shown in green as $\mathbf{N}_{\ft}$ and $\mathbf{S}_{\ft}$, respectively. The complementary region is $\mathbf{D}_{\ft}$, shown in magenta.}
        }
\label{fig:dS_Kruskal}
\end{figure}

Consider $(d+1)$-dimensional de Sitter space, dS$_{d+1}$. Recall that, topologically, dS$_{d+1}$ is $\mathbb{R} \times S_{d}$.  With de Sitter radius set equal to one, dS$_{d+1}$ is covered by ``Kruskal coordinates'' $(U, V, \Omega_{d-1})$ in terms of which the metric has the form (see e.g.~\cite{Gibbons:1977mu})
\be 
	ds^2 = - \frac{2 dU dV}{(UV - 1)^2} + \frac{(UV + 1)^2}{(UV - 1)^2} d\Omega_{d-1}^2 \ ,
\ee
where $-1 < UV < 1$. In these coordinates, future (past) infinity ($r \to \infty$ for $r$ being the radial coordinate in the static metric for dS$_{d+1}$) is reached as $UV \to 1$ with $U, V > 0$ ($U, V < 0$). The North (South) pole ($r \to 0$) of the $S_{d}$  is reached as $UV \to -1$ with $U < 0, V > 0$ ($U > 0, V < 0$). The future cosmological horizon for an observer at the North pole is the surface $U = 0, V > 0$, while the future cosmological horizon for an observer at the South pole is $U > 0, V = 0$. See figure~\ref{fig:dS_Kruskal} for an illustration.

The coordinate system naturally associated to an observer at the North or South pole of the $S_d$ is the static coordinate system $(t, r, \Omega_{d+1})$. Four patches of these coordinates are required to cover all of dS$_{d+1}$. In each patch, the metric takes on the form
\be 
	ds^2 = -(1-r^2)~ dt^2 + (1- r^2)^{-1}~ dr^2 + r^2~ d\Omega_{d-1}^2 \ .
\ee
The regions accessible to observers at the North and South poles are described by the above coordinates with $0 \leq r < 1$, with $r = 0$ being the pole and $r = 1$ the cosmological horizon. The proper time measured by either of these observers is $t$. 

\subsection{Algebras of North/South Pole Observables}
Our algebraic probe of the volume growth of de Sitter is constructed by considering the algebras of observables accessible to an observer at the North or South pole during fixed ranges of their proper time. We denote the algebra of observables accessible to an observer at the North (South) pole during the proper-time-window $t \in (\ft, \infty)$ by $\sA_{N_{\ft}}$ ($\sA_{S_{\ft}}$).\footnote{Note that we take both the proper time of the North and South observers to run forwards, so $t$ runs in the opposite direction to the static time Killing vector in the region accessible from the South pole.}

One finds that operations performed at the North pole at time $\ft$ can only effect the region of de Sitter space with $V > e^{\ft}$. Similarly, operations performed at the South pole at time $\ft$ can only effect the region of de Sitter space with $U > e^{\ft}$. By the time-like tube theorem one then has that $\sA_{N_{\ft}}$ is equivalent to algebra of all operations in the subregion $N_{\ft} = \{U < 0,~ V > e^{\ft},~ -1 < UV < 1\}$, while $\sA_{S_{\ft}}$ is equivalent to algebra of all operations in the subregion $\mathbf{S}_{\ft} = \{U > e^{\ft},~ V < 0,~ -1 < UV < 1\}$. We denote the union of these two algebras by $\sN_{\ft} \equiv \sA_{N_{\ft}} \vee \sA_{S_{\ft}}$. Assuming Haag duality for these regions, we then have that the commutant algebra $\sN'_{\ft}$ is dual to the region $\mathbf{D}_{\ft} = \{0 < U < e^{\ft},~ 0 < V < e^{\ft},~ UV < 1 \}$. See figure~\ref{fig:dS_Kruskal}.

Our probe for the volume growth of de Sitter spacetime will be the growth of the ``size'' of the algebra $\sN'_{\ft}$ as $\ft$ is increased. Note that $\sN'_{\ft}$ may be interpreted as the relative commutant for the inclusion $\sN_{\ft} \subset \sB(\sH_{dS})$, where $\sH_{dS}$ is the Hilbert space for quantum fields on the de Sitter background, suggesting that the ``size'' of $\sN'_{\ft}$ may be related to an index for this inclusion.

\subsection{Applying our proposal to de Sitter}
Our proposal is that an index for the inclusion $\sN_{\ft} \subset \sB(\sH_{dS})$ is given by the exponential of the volume of the maximal slice, $\fb_{\ft},$ of $\mathbf{D}_{\ft}$. Explicitly, our proposal is that
\be 
	\le(\sB(\sH_{dS}) : \sN_{\ft}\ri) = \exp\le(C~ \vol(\fb_{\ft})\ri) \ ,
\ee
for some constant $C$.

Note that we have the inclusion $\sN_{\ft} \subset \sN_{\ft'}$ whenever $\ft > \ft'$ which implies the inclusion $\sN'_{\ft} \supset \sN'_{\ft'}$, showing that the algebra $\sN'_{\ft}$ does indeed grow as $\ft$ is increased, reflecting the increase in $\vol(\fb_{\ft})$. Geometrically, we can think of the increase of $\ft$ as an increase in the affine time of the `cuts' of the future horizons for the North and South poles (respectively $U = 0,~ V = e^{\ft}$ and $U = e^{\ft},~ V = 0$) that define the boundary of $\mathbf{D}_{\ft}$. 

The discussion so far has been almost identical to the case of the eternal AdS black hole with the role of the Left/Right asymptotic boundary played by the South/North pole observer; however, the behavior of $\vol(\fb_{\ft})$ is dramatically different in the de Sitter case. The important difference is that, despite the similarities between the future singularity of the eternal AdS black hole and future infinity of dS$_{d+1}$ when drawn on the Penrose diagram, the future singularity is a surface with volume shrinking to zero, while future infinity of dS$_{d+1}$ is a surface of volume increasing to infinity. The implies that maxmial volume surfaces are ``repelled'' from the future singularity of the eternal AdS black hole while they are instead ``attracted'' to future infinity in dS$_{d+1}$~\cite{Susskind:2021esx}. This gives rise to different behaviors of $\vol(\fb_{\ft})$ in de Sitter versus the eternal AdS black hole, despite the fact that the algebraic probes for the ``interior'' region, $\mathbf{D}_{\ft},$ are defined in the exact same way (when we view the North/South observer as analogous to the Left/Right asymptotic boundary).

One particularly stark manifestation of this difference is that $\fb_{\ft}$ can be cutoff dependent  
in dS$_{d+1}$.
More explicitly, when $\ft > 0$, $\mathbf{D}_{\ft}$ contains points at future infinity, meaning that a maximal volume slice, $\fb_{\ft},$ of $\mathbf{D}_{\ft}$ becomes infinite for $\ft > 0$. 
We can regularize the infinity by cutting off dS at a ``large time'' by removing the part of the spacetime with $UV > 1 - \ep_{t}$ for some small $\ep_{t} > 0$. In the presence of this cutoff, the maximal volume slice of $\mathbf{D}_{\ft}$
for $\ft > 0$ runs along the cutoff hypersurface and becomes cutoff-dependent.
This is in contrast to the eternal AdS black hole case where no ``near-singularity'' cutoff was needed since maximal volume slices never approached the singularity.
The cutoff dependence here is reminiscent of the case {for an index calculation dual to}  a subregion in AdS that intersects the boundary, where the bulk volume becomes divergent and requires the introduction of an IR cutoff near the AdS boundary.

For $\ft < 0,$ the maximal volume slice, $\fb_{\ft}$, of $\mathbf{D}_{\ft}$ in de Sitter {is independent of any cutoff}; however, for $d > 1$, the computation of its volume is complicated and does not bring further insight, so we will instead content ourselves with a computation of the geodesic distance, $l_{\ft}$, between the North and South horizon `cuts' (at the same position on the $S_{d-1}$) that define $\mathbf{D}_{\ft}$.\footnote{No {smooth and cutoff-independent} geodesic connects these `cuts' for $\ft > 0$.} We will call $l_{\ft}$ the `length' of $\mathbf{D}_{\ft}$ and view it as a proxy for $\vol(\fb_{\ft})$. The geodesic distance, $l_{\ft}$, from $(U = e^{\ft}, V = 0, \Omega_{d-1})$ to $(U = 0, V = e^{\ft}, \Omega_{d-1})$ is easily computed by viewing dS$_{d+1}$ as a hypersurface in $\mathbb{R}^{1,d+1}$. The result is
\be 
	l_{\ft} = \arccos\le(1 - 2 e^{2\ft} \ri), \qquad \ft < 0 \ .
\ee

The `length' of $\mathbf{D}_{\ft}$ has the following early time behavior
\be 
	l_{\ft} = 2 e^{\ft} + \frac{e^{3\ft}}{3} + O(e^{5\ft}),~ \ft \to -\infty \ .
\ee
We see that the length of $\mathbf{D}_{\ft}$ grows exponentially with the observers' early proper time cutoff $\ft$ as $\ft \to - \infty$.

The `length' of $\mathbf{D}_{\ft}$ has the following near-zero time behavior
\be 
	l_{\ft} = \pi - 2\sqrt{2} \sqrt{-t} + O((-t)^{3 \ov 2}),~ \ft \to 0^- \ .
\ee
Thus the `length' of $\mathbf{D}_{\ft}$ goes to a constant as $\ft$ goes to zero from below (and $\mathbf{D}_{\ft}$ approaches future infinity). For $\ft > 0$, the `length' of $\mathbf{D}_{\ft}$ {diverges as the ``large-time'' cutoff is removed}.

{With this behavior of the `length' of $\mathbf{D}_{\ft}$ understood, we see that~\eqref{v1} implies that there is discontinuous jump in the embedding of $\sN_{\ft}$ into $\sB(\sH_{\rm dS})$ as $\ft$ crosses zero. In particular, for $\ft < 0$, our proposal suggests that the number of operators from $\sB(\sH_{\rm dS})$ that {are not contained in} $\sN_{\ft}$ is independent of any ``large time'' cutoff, whereas for $\ft > 0$ this number of operators becomes cutoff-dependent. It is interesting to note that this transition precisely occurs where signals sent from $\sA_{S_{\ft}}$ and $\sA_{N_{\ft}}$ can no longer interact at any point in the future.}

{The divergence in the volume of maximal slices in dS has been interpreted as an exponential rate of complexity growth at late times~\cite{Susskind:2021esx} and has been called ``hyperfast'' complexity growth. It would be interesting to understand if our proposed discontinuous jump in algebra ``size'' can be connected to ``hyperfast'' complexity growth.}

\section{Discussion} \label{sec:disc}

{In this paper, we have proposed that the maximal volume slice of a bulk domain of dependence defines a new type of index for a corresponding boundary algebra inclusion. Heuristically, the exponential of the bulk volume measures the ``size'' of a relative commutant algebra on the boundary which characterizes the boundary algebra inclusion. We have discussed how this index shares {some} properties with the standard Kosaki index~\cite{kosakiIndex}, but is also not equivalent to this standard definition. We have also computed this holographic index in many cases of interest.}

{Applying our proposal to the Wheeler-de Witt patch of an eternal AdS black hole we have argued  that the shrinking of a particular boundary algebra with the increase of the left and right boundary anchoring time is dual to the bulk volume growth. We then described a heuristic connection between algebra shrinking and boundary complexity growth. We discussed two possible interpretations of our index formula and argued that the well-definedness of the generalized entropy at finite $G_N$ suggests that our index can be extended to the finite $N$ setting; however, we do not have any explicit description of this extension.} {Finally, we applied our proposal to two antipodal observers in de Sitter space and discussed the algebraic consequences implied by the rapid growth of the volume of the region between the two observers' future cosmological horizons.}

{Our proposal provides important insights into the boundary characterization of bulk subregions in holographic systems; however, there are still many aspects that should be better understood. The most pressing question raised by our proposal is: can we understand physics in bulk subregions at finite $G_N$ through state-dependent type I boundary algebras? We have argued that such a description should exist and our proposal gives a conjecture for the size of these algebras, however an explicit description remains elusive.}

\vspace{0.2in}   \centerline{\bf{Acknowledgements}} \vspace{0.2in}
We would like to thank Netta Engelhardt, Elliott Gesteau, 
Jonah Kudler-Flam,
Juan Maldacena, Mudassir Moosa, and Gautam Satishchandran
for discussions. 
This work is supported by the Office of High Energy Physics of U.S. Department of Energy under grant Contract Number  DE-SC0012567 and DE-SC0020360 (MIT contract \# 578218).
S.L. is supported by the Martin A.~and Helen Chooljian Member Fund and the Fund for Natural Sciences at the Institute for Advanced Study and the National Science Foundation under the grant PHY-2207584.

\appendix

\section{Index of Inclusions of Algebras} \label{app:index}

In this section we review some aspects of the theory of index of inclusions of von Neumann algebras. 
We first discuss Kosaki's index of inclusion~\cite{kosakiIndex}, which applies to general inclusions of von Neumann factors, and show in detail how this reproduces our heuristically defined index for finite type I inclusions. Kosaski's definition can be applied to an arbitrary inclusion of von Neumann algebras and agrees with the definition of the Jones index for type II$_1$ factors~\cite{jonesIndex} as well as a definition of index for inclusions of type III$_1$ factors given by Longo~\cite{longoIndex}.

\subsection{Kosaki's index of inclusion}

An index for inclusions of type II$_1$ subfactors was introduced by Jones~\cite{jonesIndex}.
Here we describe a general definition of the index for an inclusion for factors due to Kosaki~\cite{kosakiIndex}, which applies to arbitrary factors and thereby generalizes the 
Jones index to type III algebras.  We will specialize to the case of standard representations as this is the most relevant situation for physics, but refer readers to~\cite{kosakiIndex} for the more general definition.
We will also discuss indices of Jones and Longo which are more directly applicable to some physically relevant scenarios. 

Let $\sN \subset \sM$ be an inclusion of von Neumann factors acting on a common Hilbert space $\sH_{\rm tot}$ that provides a standard representation for $\sM.$ Assume that there exists a conditional expectation $E : \sM \to \sN$.\footnote{A  conditional expectation is a positive, linear, unital map $E$ between a von Neumann algebra $\sM$ and a subalgebra $\sN$,  $E : \sM \to \sN$ satisfying: $E(n) =n$ for $n \in \sN$ and $E(n_1mn_2) = n_1E(m)n_2$ where $m \in \sM$ and $n_{1,2} \in \sN$.}  Suppose we are given two states $\ket{\phi},~\ket{\psi} \in \sH_{\rm tot}$ and that $\ket{\phi}$ is cyclic and separating for $\sM.$ We will use the (non-ket) notation $\phi,~\psi : \sM \to \mathbb{C}$ to denote the algebraic states induced by the vectors $\ket{\phi},\ket{\psi}.$ An ``inverse'' of the state $\phi$ can be obtained as an operator-valued weight $\phi^{-1} : \sB(\sH_{\rm tot}) \to \sM'$ such that
\be \label{eq:defInvState}
	\bra{\psi} \phi^{-1}(a) \ket{\psi} = \Tr\le( \De_{\psi | \phi} a \ri), \qquad a \in \sB(\sH_{\rm tot}) \ ,
\ee
where $\De_{\psi | \phi}$ is the relative modular operator from $\phi$ to $\psi$ with respect to $\sM$ and $\Tr$ denotes the Hilbert space trace on $\sH_{\rm tot}$. Despite the appearance of the state $\ket{\psi}$ in the definition~\eqref{eq:defInvState}, the operator-valued weight denoted by $\phi^{-1}$ can be shown to be independent of $\ket{\psi}$ and only depends on $\ket{\phi}.$ 

{To understand better} why the notation $\phi^{-1}$ makes sense (and how it is independent of $\psi$ despite appearances), consider the case of the standard representation of a type I${}_m$ factor, i.e. write $\sM = \sB(\sH_m) \otimes \mathbf{1}_{\bar{\sH}_m}.$ Since $\ket{\phi}$ is cyclic and separating for $\sM,$ it can be described on $\sM$ by a full-rank reduced density matrix $\rho_{\phi} \in \sM.$ The relative modular operator can be computed to be
\be 
	\De_{\psi|\phi} = \rho_{\phi}^{-1} \otimes \rho'_{\psi} \ ,
\ee
where $\rho'_{\psi} \in \sM'$ is the reduced density matrix describing $\psi$ on $\sM'$. In this case~\eqref{eq:defInvState} is
\be 
	\bra{\psi}\phi^{-1}(a)\ket{\psi} = {\rm Tr}_{\bar{\sH}_m}\le(\rho'_{\psi} \phi^{-1}(a)\ri) = {\rm Tr}_{\sH_m \otimes \bar{\sH}_m}\le((\rho_{\phi}^{-1} \otimes \rho'_{\psi})a \ri) \ ,
\ee
and therefore,
\be \label{eq:finDimInvState}
	\phi^{-1}(a) = {\rm Tr}_{\sH_m}\le(\rho_{\phi}^{-1} a\ri) \ .
\ee
This is completely determined by $\phi$ and is the `inverse' of the state in the sense that $\rho_{\phi}^{-1}$ appears.

Similar manipulations allow us to define the notion of `inverse' for general operator-valued weights. We will be particularly interested in the inverse of the conditional expectation $E.$ The `inverse' of a conditional expectation is a map on the commutants, i.e. $E^{-1} : \sN' \to \sM'.$ In the general setting, it is defined through the relation of maps from $\sB(\sH_{\rm tot})$ to $\sN$:
\be \label{eq:defInvCondExp}
	\le(\psi' \circ E^{-1}\ri)^{-1} = E \circ \psi'^{-1} \ ,
\ee
for any weight $\psi' : \sM' \to \mathbb{C}$ on $\sM'$. Despite the apparent dependence on $\psi',$ it can be shown that $E^{-1}$ is completely determined by $E.$ It can also be shown that $E^{-1}(\mathbf{1}_{\sN'})$ is proportional to $\mathbf{1}_{\sM'},$ the identity in $\sM'$, so long as $\sN$ and $\sM$ are factors. We refer the reader to~\cite{kosakiIndex} for details. In this very general setting, Kosaki's index for the conditional expectation $E : \sM \to \sN$, ${\rm Ind} E$, is then defined by
\be \label{eq:defKosakiIndex}
	E^{-1}\le(\mathbf{1}_{\sN'}\ri) = {\rm Ind} E \cdot \mathbf{1}_{\sM'} \ .
\ee

We emphasize that this index is defined for any conditional expectation from $\sM$ to $\sN$ no matter which type of factors they are since equations~\eqref{eq:defInvState} and~\eqref{eq:defInvCondExp} are defined in complete generality.\footnote{In fact, using Connes' theory of spatial derivatives~\eqref{eq:defInvState} can be generalized further to the case of a non-standard representation of $\sM$; however such a generalization is not needed for this paper so we omit its description.}

{As an example, we consider} an inclusion of finite dimensional type I factors, i.e. $\sN =\mathbf{1}_k \otimes  \sB(\sH_n)  \subset \sM = \sB (\sH_m) =  \sB(\sH_n) \otimes \sB(\sH_k),$ where $\sH_{n,m,k}$ are Hilbert spaces of dimension $n,m,k,$ respectively. A conditional expectation of interest is the normalized partial trace 
\be \label{eq:finDimCondExp}
	E = \frac{1}{k} {\rm tr}_{\sH_k} : \; \sB(\sH_n) \otimes \sB(\sH_k) \to \sB(\sH_n) \ .
\ee
{Now consider} the `doubled' Hilbert space $\sH_{\rm ext} = \sH_m \otimes \overline{\sH_m}$, which furnishes a standard representation of $\sM,$ i.e. a representation that contains a cyclic and separating vector for $\sM$. 
On $\sH_{\rm ext}$, $\sN' = \mathbf{1}_{\sH_n} \otimes \sB(\sH_k) \otimes \sB(\overline{\sH_m})$ and $\sM' = \mathbf{1}_{\sH_m} \otimes \sB(\overline{\sH_m})$. We can now explicitly compute the inverse, $E^{-1}:\sN' \to \sM'$, of the normalized partial trace~\eqref{eq:finDimCondExp}. Using~\eqref{eq:finDimInvState} one easily sees that for $a \in \sB\le(\sH_{\rm tot}\ri) = \sB(\sH_n \otimes \sH_k \otimes \bar{\sH}_m)$ we have
\be 
\begin{aligned}
	E \circ \psi'^{-1}(a) &= \frac{1}{k} \tr_{\sH_k}\le(\tr_{\bar{\sH}_m} \le(\rho'^{-1}_{\psi} \cdot a\ri)\ri) \\
	&= \tr_{\sH_k \otimes \bar{\sH}_m}\le(\rho'^{-1}_{\psi \circ E^{-1}} \cdot a\ri) \ .
\end{aligned}
\ee
From this we can easily read off for $n' \in \sN'$
\be 
	\psi' \circ E^{-1} (n') = \tr_{\sH_k \otimes \bar{\sH}_m}\le(\le(k \mathbf{1}_{\sH_k} \otimes \rho'_{\psi} \ri) \cdot n'\ri) = \tr_{\bar{\sH}_m}\le(\rho'_{\psi} \cdot k \tr_{\sH_k} \le(n'\ri)\ri) \ ,
\ee
so we see that
\be 
	E^{-1}(n') = k \tr_{\sH_k}\le(n'\ri) \ .
\ee
Clearly $E^{-1}(\mathbf{1}_{\sH_k \otimes \bar{\sH}_m})$ is proportional to $\mathbf{1}_{\bar{\sH}_m}$ with proportionality coefficient $k^2.$ The index is defined as the proportionality coefficient so we find 
\be 
	{\rm Ind} E = k^2 \ ,
\ee
exactly as in our ``counting copies'' of $\sN$ in $\sM$ argument in the introduction.

\subsubsection{The Jones index}
The Jones index~\cite{jonesIndex} characterizes an inclusion of type II$_1$ factors, $\sN \subset \sM$. We review a simple characterization of the Jones index here.

Consider $\sN \subset \sM$ represented on the GNS Hilbert space, $\sH_{\sM} = L^2(\sM),$ of the (unique) normalized tracial state, $\tr_{\sM}: \sM \to \mathbb{C}$, on $\sM$. This is a standard representation of $\sM$ with the representative of the trace (GNS vacuum), $\ket{\Om} \in \sH_{\sM}$ such that $\bra{\Om} \cdot \ket{\Om} = \tr_{\sM}(\cdot)$, being cyclic and separating. However, because $\sN$ is a proper subalgebra of $\sM$, $\ket{\Omega}$ may not be cyclic for $\sN$. Instead, the action of $\sN$ on $\ket{\Omega}$ may only span a proper subspace of $\sH_{\sM}$. We denote the closure of this subspace by $V_{\sN} \equiv \overline{\pi_{\tr}(\sN) \ket{\Omega}} \subset \sH_{\sM}$. The orthogonal projection, $e_{\sN},$ onto the subspace $V_{\sN}$ is called the Jones projection. It is an element of $\sN'$ and the Jones index can be defined by
\be 
	\le[ \sM : \sN \ri] = \frac{1}{\tr_{\sN'}(e_{\sN})} \ .
\ee

One can show that Kosaki's index agrees with that of Jones when applied to the conditional expectation induced by the tracial state for an inclusion of type II$_1$ algebras.

For $a, b \in \pi_{\tr}(\sM) \subset \sB(\sH_{\sM}),$ since $\tr_{\sM}$ is a trace (i.e. has the cyclic property) we have
\be 
	\bra{\Om} a b \ket{\Om} = \bra{\Om} b a \ket{\Om} \Rightarrow \De_{\Om} = \mathbf{1}_{\sH_{\sM}} \ .
\ee	
Restricting to operators in $\sM'$ one has that $\bra{\Om} m' \ket{\Om} = \tr_{\sM'}(m'),~ \forall m' \in \sM',$ with $\tr_{\sM'}$ the normalized trace on $\sM'.$ 

For any weights $\psi': \sM' \to \mathbb{C}$ and $\phi: \sN \to \mathbb{C}$ and normal faithful conditional expectation $E: \sM \to \sN$ we have $\De_{\psi' \circ E^{-1} ~|~ \phi} = \De_{\psi' ~|~ \phi \circ E}$. In this case we are interested in $E$ being the conditional expectation induced by the trace on $\sM,$ $\psi' = \tr_{\sM'},$ and $\phi = \tr_{\sN}.$ For this special choice of conditional expectation we have (using $e_{\sN}\ket{\Om} = \ket{\Om}$ several times)
\be 
\begin{aligned}
	{\rm tr}_{\sM}(ab) &= \bra{\Om} e_{\sN} ab e_{\sN} \ket{\Om} = \bra{\Om} E(ab) \ket{\Om} = \bra{\Om} E(ba) \ket{\Om} \propto {\rm tr}_{\sN} \le(E (ab) \ri)\ .
\end{aligned}
\ee
Note that, restricted to $\sN$, $\bra{\Om} \cdot \ket{\Om},$ is still cyclic in its argument so it must be proportional to the (normalized) trace on $\sN.$ Taking $ab = \mathbb{1}_{\sH_{\sM}} \in \pi_{tr}(\sN)$ we have that $E(\mathbb{1}_{\sH_{\sM}}) = \mathbb{1}_{\sH_{\sM}}$ and thus
\be 
\begin{aligned}
	{\rm tr}_{\sM}(\mathbf{1}_{\sM}) = \langle{\Om|\Om}\rangle = 1 \propto {\rm tr}_{\sN}(\mathbf{1}_{\sN}) \ ,
\end{aligned}
\ee
but ${\rm tr}_{\sN}(\mathbf{1}_{\sN}) = 1$ by definition of normalized trace so the proportionality constant is $1$ and we have ${\rm tr}_{\sM} = {\rm tr}_{\sN} \circ E.$ Thus, we have the following equality of spatial derivatives $\De_{\tr_{\sM'} \circ E^{-1} ~|~ \tr_{\sN}} = \De_{\tr_{\sM'} ~|~ \tr_{\sN} \circ E} = \De_{\tr_{\sM'} ~|~ \tr_{\sM}}$, but this is equal to $\De_{\Om ~|~ \Om} = \De_{\Om} = \mathbb{1}_{\sH_{\sM}}$ since $\ket{\Om}$ is the representative of both the tracial state on $\sM$ and that on $\sM'.$ We therefore have $\De_{\tr_{\sM'} \circ E^{-1} ~|~ \tr_{\sN}} = 1$. We must therefore have that $\tr_{\sM'} \circ E^{-1}$ is proportional to the trace on $\sN',$ i.e.
\be 
	{\rm tr}_{\sN'}(\cdot) = \lambda~ {\rm tr}_{\sM'} \circ E^{-1}(\cdot) \ , 
\label{eq:propNpMpTraces}
\ee 
for some proportionality constant $\lambda.$ Recalling that Kosaki's index is defined by $E^{-1}(\mathbf{1}_{\sN'}) = {\rm Ind}E~\mathbf{1}_{\sM'}$ so we have
\be 
	{\rm tr}_{\sN'}(\mathbf{1}_{\sN'}) = \lambda~{\rm Ind}E~ {\rm tr}_{\sM'} (\mathbf{1}_{\sM'}) \ .
\ee
The traces are defined to each be normalized so we must have $\lambda~{\rm Ind}E = 1$.

To compute $\lambda$ consider inserting $e_{\sN} \in \sN'$ into~\eqref{eq:propNpMpTraces}. One obtains
\be 
	{\rm tr}_{\sN'}(e_{\sN}) = \lambda~ {\rm tr}_{\sM'} \circ E^{-1}(e_{\sN}) = \lambda \ .
\ee

The Jones index is given by
\be 
	[\sM : \sN] = \frac{1}{{\rm tr}_{\sN'}\le(e_{\sN}\ri)} \ ,
\ee
which is clearly $\lambda^{-1}$ so $\lambda {\rm Ind}E = 1$ implies
\be 
	{\rm Ind}E = [\sM : \sN] \ ,
\ee
proving equality of the Kosaki index and Jones index in this case.

\subsubsection{Longo's index}
Longo's index~\cite{longoIndex} characterizes an inclusion of properly infinite factors. In particular, it applies to inclusions of type III$_1$ factors. We briefly describe its definition below.

Consider an inclusion of type III$_1$ factors $\sN \subset \sM$ acting on a common Hilbert space $\sH$ that is a standard representation for both $\sN$ and $\sM$, with common cyclic and separating vectors $\ket{\Om} \in \sH$. 
Assume that we have a normal conditional expectation $E : \sM \to \sN$.\footnote{Note that this assumption fails for many interesting inclusions of type III$_1$ factors in relativistic QFT.} We then have, for any faithful state $\phi : \sN \to \mathbb{C}$ on $\sN$, that the modular automorphism groups of the state $\phi$ on $\sN$ and that of $\phi \circ E$ on $\sM$ are identical when restricted to $\sN$. This implies that appropriate modular crossed products of $\sN$ and $\sM$ also satisfy containment.\footnote{Note that modular crossed products do not preseve containment in general. The existence of the normal conditional expectation $E$ facilitates the containment in this very special setting.} Explicitly, denoting the modular automorphism group of $\sN$ with respect to $\phi$ by $\sigma^{\phi}$ and that of $\sM$ with respect to $\phi \circ E$ by $\sigma^{\phi \circ E}$, we have an inclusion of type II$_{\infty}$ factors
\be 
	\widetilde{\sN}  = \sN \rtimes_{\sigma^{\phi}} \mathbb{R} \subset \widetilde{\sM}  = \sM \rtimes_{\sigma^{\phi \circ E}} \mathbb{R} \ .
\label{eq:containedCPs}
\ee

Longo's index concerns the transformation of the trace $\tr_{\widetilde{\sM}}$ under a particular endomorphism of $\widetilde{\sM}$. Denote the modular of conjugations of $\widetilde{\sN}$ and $\widetilde{\sM}$ with respect to their dual weights by $J_{\widehat{\phi}}$ and $J_{\widehat{\phi \circ E}}$ respectively. The ``canonical endomorphism'' from $\widetilde{\sM}$ to $\widetilde{\sN}$ is defined as
\be 
	\widetilde{\gamma}(\widetilde{m}) = J_{\widehat{\phi}} J_{\widehat{\phi \circ E}} \widetilde{m} J_{\widehat{\phi \circ E}} J_{\widehat{\phi}}, \quad \forall \widetilde{m} \in \widetilde{\sM} \ .
\ee
One can show that the functional $\tr_{\widetilde{\sM}} \circ \widetilde{\gamma}$ on $\widetilde{\sM}$ must be proportional to $\tr_{\widetilde{\sM}}$ itself. Longo's index is defined as this proportionality coefficient, i.e.
\be 
	{\rm Ind}E \equiv \lambda \quad {\rm s.t.} \quad {\rm tr}_{\widetilde{\sM}} \circ \widetilde{\gamma} = \lambda \cdot {\rm tr}_{\widetilde{\sM}} \ .
\label{eq:longoIndexDef}
\ee

\section{Useful results on half-sided modular translations} \label{app:HSMT} 

\subsection{Half-sided modular translations in the large dimension limit} \label{app:HSMTa}

In this appendix we discuss the geometric transformations generated by $P_+,~P_-,$ and $P_+ + P_-$ in the limit of large mass of the bulk field (large boundary scaling dimension) propagating on the eternal AdS$_{d+1}$-Schwarzschild geometry. The results below hold only when we restrict to modes of the bulk field with zero spatial momentum and apply in the limit $m^2 l_{\rm AdS}^2 >> 1$.\footnote{These results hold in general in AdS$_2$ since there is then no notion of transverse spatial momentum at all. The result for the action of $\exp(is(P_- + P_+))$ can be proven by combining the results of~\cite{longPaper} and those of~\cite{Lashkari:2024lkt}.}

In Kruskal coordinates $\exp(iuP_+)$ implements the bulk pointwise transformation $\phi_{\vk  = 0}(U, V) \to \lambda_u~\phi_{\vk  = 0}(U_u, V_u)$ with 
\be 
	U_u = U + u, \qquad V_u = \frac{V}{1 - u V} \ .
\ee
The corresponding action on a left/right boundary GFF operator (of zero spatial momentum) at time $t$ is $\sO_{L/R}(t) \to \lambda_u \sO_{L/R}(f_{L/R}^+(t; u))$ with 
\be
	f_L^+(t; u) = t + \log(1 + u e^{-t}), \quad f_R^+(t; u) = t - \log(1 - u e^{t}) \ ,
\ee
for all values of $u$ for which $f_{L/R}^+(t; u)$ is real.

Similarly $\exp(ivP_-)$ implements the bulk pointwise transformation $\phi_{\vk  = 0}(U, V) \to \lambda_v~\phi_{\vk  = 0}(U_v, V_v)$ with 
\be 
	U_u = \frac{U}{1 - v U}, \qquad V_u = V + v \ .
\label{eq:bulkPXfm}
\ee
The corresponding action on a left/right boundary GFF operator (of zero spatial momentum) at time $t$ is $\sO_{L/R}(t) \to \lambda_v \sO_{L/R}(f_{L/R}^-(t; v))$ with 
\be
	f_L^-(t; v) = t - \log(1 - v e^{t}), \quad f_R^-(t; v) = t + \log(1 + v e^{-t}) \ ,
\ee
for all values of $v$ for which $f_{L/R}^-(t; v)$ is real.

Finally $\exp(is(P_+ + P_-))$ implements the bulk pointwise transformation $\phi_{\vk  = 0}(U, V) \to \lambda_s~\phi_{\vk  = 0}(U_s, V_s)$ with 
\be 
	U_s = \frac{U\cos(s) + \sin(s)}{\cos(s) - \sin(s) U}, \qquad V_s = \frac{V\cos(s) + \sin(s)}{\cos(s) - \sin(s) V} \ .
\label{eq:bulkPMXfm}
\ee
The corresponding action on a left/right boundary GFF operator (of zero spatial momentum) at time $t$ is $\sO_{L/R}(t) \to \lambda_s \sO_{L/R}(f(t; s))$ with 
\be
	f(t; s) = t - \log\le[ \frac{\cos s - \sin s e^t}{\cos s + \sin s e^{-t}} \ri] \ ,
\ee
for all values of $s$ for which $f(t; s)$ is real.

\subsection{AdS$_2$ calculations} \label{app:HSMTb} 
We now discuss the special case of our procedure using half-sided modular translations to unitarily map time-band algebras to other time-band algebras in the special case of AdS$_2$ where everything may be computed explicitly. We first compute the volume of the maximal slice of the Wheeler-de Witt patch (which is just the (regularized) length of the AdS$_2$ wormhole in this case) and then show that unitary evolution with half-sided modular inclusion can be used to compute volume changes entirely on the $t=0$ slice.

\subsubsection{Regulated AdS$_2$ `wormhole' length}
In AdS$_2$ the volume of the maximal slice of the Wheeler-de Witt (WdW) patch is simply the distance between the two asymptotic boundaries. The formula for the length of the wormhole at fixed boundary times $t_R = t_L = \ft$ has then very simple analytic expression (unlike formulas for the volume of the WdW patch in higher dimensions). 

There is no eternal AdS black hole in two dimensions; however, there is a configuration with similar causal structure. This is when we consider the ``Rindler patches'' of global AdS$_2$. Global AdS$_2$ is described in global $(\tau, \sigma)$ coordinates by metric
\be 
	ds^2 = \frac{l^2}{\sin^2\sigma} \le(-d\tau^2 + d\sigma^2\ri) \ ,
\ee
where $l$ is the AdS length scale and we have $\tau \in \mathbb{R},~ \sigma \in (0,\pi).$ The left asymptotic boundary is at $\sigma = 0$ and right asymptotic boundary is at $\sigma = \pi.$

AdS$_2$ can also be described in ``Schwarzschild'' coordinates $(t, r)$ with metric
\be 
	ds^2 = - f(r) dt^2 + \frac{dr^2}{f(r)}, \qquad f(r) = \frac{r^2 - r_H^2}{l^2} \ ,
\label{eq:AdS2SchwMet}
\ee
for $t \in \mathbb{R},~ r \in (r_H, \infty),$ where $r_H > 0$ is the horizon `radius' (related to the Hawking temperature w.r.t $t$). This is precisely the functional form of the Schwarzschild metric so these coordinates are referred to as Schwarzchild coordinates. These coordinates only cover patches of global AdS$_2$ and the total of the four patches for which the metric has the form~\eqref{eq:AdS2SchwMet} has the qualitative causal structure features analogous to those of higher-dimensional eternal AdS black holes. We are most interested in the `right exterior' patch with $\tau - \sigma \in (-3\pi/2, -\pi/2),~ \tau + \sigma \in (\pi/2, 3\pi/2),~ \sigma \in (0,\pi)$. In this region, the mapping between Schwarzschild and global coordinates is given by
\be 
\begin{aligned}
	\tau &= \arctan\le(e^{\frac{r_H}{l^2} t} \sqrt{r - r_H \ov r + r_H}\ri) + \arctan\le(-e^{-\frac{r_H}{l^2} t} \sqrt{r - r_H \ov r + r_H}\ri) \\
	\sigma &= \arctan\le(e^{\frac{r_H}{l^2} t} \sqrt{r - r_H \ov r + r_H}\ri) - \arctan\le(-e^{-\frac{r_H}{l^2} t} \sqrt{r - r_H \ov r + r_H}\ri) + {\pi \ov 2} \ .
\end{aligned}
\ee

We note that it is only the portion of AdS$_2$ covered by Schwarzschild coordinates that we expect to be described holographically by two quantum mechanical systems entangled in the TFD. The temperature of~\eqref{eq:AdS2SchwMet} with respect to $t$ is the temperature of the TFD in which the two quantum mechanical systems are entangled.

In global coordinates, the geodesic connecting left and right boundaries at asymptotic global time $\tau_0$ is simply the constant $\tau = \tau_0$ slice of the bulk. Without IR cutoffs such a length is clearly infinite and $\tau_0$-independent. However, for the analogy to higher-dimensional black hole physics we must consider only the Schwarzschild portion of AdS$_2$. The natural bulk IR cutoff in such coordinates is at $r = r_{\infty} >> r_H$. This is a $\tau$-dependent timelike surface in AdS$_2$ that breaks the $\tau$-translation symmetry and thereby allows for time-dependence in the wormhole length. On this surface the global time, $\tau$, and Schwarzschild time, $t$, are approximately related by 
\be 
	\tan(\tau) = \sinh\le(\frac{r_H}{l^2} t\ri) + O\le(\le(\frac{r_H}{r_{\infty}}\ri)^2\ri) \ .
\label{eq:AdS2GlobToSchwTimeBdry}
\ee

At fixed Schwarzschild time $t = \ft$ on the $r = r_{\infty}$ surface, the global radial coordinate is given by
\be
	 \sigma|_{r=r_{\infty}} = \pi - \frac{r_H}{r_{\infty} \cosh\le(\frac{r_H}{l^2} \ft\ri)} + O\le(\le(\frac{r_H}{r_{\infty}}\ri)^2\ri) \ ,
\ee
which using~\eqref{eq:AdS2GlobToSchwTimeBdry} can be described in global coordinates as the surface
\be 
	\sigma|_{r=r_{\infty}} = \pi - \frac{r_H}{r_{\infty}} \cos(\tau) + O\le(\le(\frac{r_H}{r_{\infty}}\ri)^2\ri) \ ,
\ee
for $\tau \in (-\pi/2, \pi/2).$

At fixed boundary (Schwarzschild) time $t_L = t_R = \ft$ we then have that the regularized length of the wormhole is (by reflection symmetry about $\sigma = \pi/2$) 
\be 
\begin{aligned}
	{\rm length} &= 2 \int_{\pi/2}^{\sigma|_{r=r_{\infty}}} \frac{l}{\sin \sigma} d\sigma = -2l \arctanh(\cos(\sigma))|_{\sigma = \pi/2}^{\sigma|_{r=r_{\infty}}} \\
	&= 2l \le(\log\le(\frac{2 r_{\infty}}{r_H}\ri) + \log\cosh \le(\frac{r_H}{l^2} \ft\ri)\ri) + O\le(\le(\frac{r_H}{r_{\infty}}\ri)^2\ri) \ .
\end{aligned}
\label{eq:AdS2WHLength}
\ee

\subsubsection{Evolution with HSMTs}
We now show that evolution with operators obtained from half-sided modular translations (HSMTs) can be used to map the WdW patch at $t_L = t_R = \ft$ back to a new WdW patch at $t_L = t_R = 0.$ One finds that the length of the new WdW patch centered on the $t=0$ slice has precisely the same volume as the original WdW patch `centered' at $t=\ft$ but we now have an algebra inclusion of that of the original WdW patch centered at $t=0$ (commutant of the boundary algebra associated to the union of left and right time bands each of width $2w$) in that of the new WdW patch.

The evolution implemented by operators constructed from HSMTs is most easily described in Kruskal coordinates $(U,V)$ which cover the same portion of global AdS$_2$ as the four regions of Schwarzschild coordinates. In Kruskal coordinates, the AdS$_2$ metric is
\be 
	ds^2 = -\frac{4l^2}{(1+UV)^2}~ dU~dV \ ,
\ee
and we have $-1 < UV < 1$ with $UV \to -1$ the asymptotic boundaries and $UV \to 1$ the ``singularity'' ($r \to 0$).

In the large mass limit of a bulk scalar field propagating on AdS$_2$ there is a one-parameter family of unitary operators $\exp(i(P_+ + P_-)s)$ that implements a pointwise transformation on the bulk field $\phi(U,V) \to \lambda_s \cdot \phi(U_s, V_s)$ with
\be 
	U_s = \frac{U \cos(s) + \sin(s)}{\cos(s) - U\sin(s)}, \qquad V_s = \frac{V \cos(s) + \sin(s)}{\cos(s) - V\sin(s)} \ .
\ee

Using $\exp(i(P_+ + P_-)s)$ for a special value $s=s_*,$ this transformation maps left/right equal-width boundary time-bands centered at some $t=\ft$ back to new equal-width time-bands centered at $t=0$. Explicitly, in terms of dimensionless boundary time $\eta = (r_H/l^2) t$ this transformation maps a boundary time-interval $\eta \in (\eta_0 - w_{\eta}, \eta_0 + w_{\eta})$ to a new time-interval $\eta \in (-\eta_*, \eta_*)$ with
\be 
	\eta_* = - \log\le(\frac{\sqrt{(\cosh(2\eta_0) + \cosh(2w_{\eta}))/2} - \sinh(w_{\eta})}{\cosh(\eta_0)}\ri) \ .
\ee
In the limit of very small $w_{\eta} << 1$ the commutant of the algebra of the union of these left and right time-bands describe (regularized) WdW patches in the bulk. We are interested in comparing the length of the dual of the commutant of the union of left/right time-band algebras for $\eta \in (-w_{\eta}, w_{\eta})$ and for $\eta \in (-\eta_*, \eta_*)$.

In the limit $w_{\eta} \to 0^+$, one also has $\eta_* \to 0^+$ so the dual of each of the commutant of each of these time-band algebras reaches out to a very large value of the Schwarzschild radial coordinate $r(w_{\eta}) >> r_H$ and $r(\eta_*) >> r_H$ respectively. The length of the corresponding duals may then be computed by setting $\ft = 0$ and replacing $r_{\infty} \to r(w_{\eta})$ or $r_{\infty} \to r(\eta_*)$ in~\eqref{eq:AdS2WHLength}.

For a time-band of dimensionful temporal half-width $w_t$ centered on $t = 0,$ the minimal radial coordinate reached in the dual is found to be
\be 
	r(w_t) = \frac{r_H}{\tanh\le(\frac{r_H}{l^2} w_t\ri)} \to_{w_t \to 0^+} \frac{l^2}{w_t} \ ,
\ee
which is clearly much larger than $r_H$ for $r_H w_t << l^2$.

We therefore find that
\be 
	r(w_t) = \frac{l^2}{w_t}, \qquad r(w_*) = \frac{l^2}{w_t}\cosh\le(\frac{r_H}{l^2} \ft\ri)
\ee
for $r_H w_t << l^2$ and thus the difference in lengths computed by~\eqref{eq:AdS2WHLength} for the evolved and un-evolved time-bands centered at $t=0$ precisely reproduces the growth in the volume of the maximal slice of the WdW patch.

Denoting the algebra for the union of left and right time-bands with $t \in (\ft-w,\ft+w)$ as $\sN_{\ft, w}$, we have therefore shown that 
\be 
	{\vol}({\rm dual}((\sN_{0, w_*})')) - {\vol}({\rm dual}((\sN_{0, w_t})')) = 2l \log\cosh\le(\frac{r_H}{l^2} \ft \ri) + O\le(\le(\frac{r_H w_t}{l^2}\ri)^2\ri) \ .
\ee
Notice that this difference of volumes is IR finite in the bulk.

We have that $(\sN_{0, w_*})'$ is unitarily equivalent to $(\sN_{\ft, w_t})'$ and we found that these algebra indeed have the same ``size'', in the sense that the maximal slices of their bulk duals have equal volume.\footnote{Note that this had to be the case since $\exp(i(P_++P_-)s)$ implements an isometry in AdS$_2$.}

\end{document}